\documentclass[fleqn,usenatbib]{mnras}
\usepackage{newtxtext,newtxmath}
\usepackage[T1]{fontenc}

\DeclareRobustCommand{\VAN}[3]{#2}
\let\VANthebibliography\thebibliography
\def\thebibliography{\DeclareRobustCommand{\VAN}[3]{##3}\VANthebibliography}

\usepackage{graphicx,psfrag}
\usepackage{mathrsfs}
\usepackage{amsmath}
\usepackage{amsfonts}
\usepackage{multirow}
\usepackage{comment}
\usepackage{ulem}
\usepackage{multirow}
\usepackage{hyperref}

\usepackage{pifont} 
\newcommand{\be}{\begin{equation}}
\newcommand{\ee}{\end{equation}}
\newcommand{\bea}{\begin{eqnarray}}
\newcommand{\eea}{\end{eqnarray}}
\newcommand{\bel}{\begin{align}}
\newcommand{\eel}{\end{align}}

\def\lm{{\ell m}}

\def\M{{\cal M}}

\def\ergsec{{\rm erg/s^{}}}
\def\ergsecg{{\rm erg/g/s}}
\def\ergseccm3{{\rm erg/cm^{3}/s}}
\def\gccm{{\rm g/cm^{3}}}
\def\Msun{{\rm M_{\odot}}}
\def\Mo{{\Msun}}

\def\GMc2{{\rm G M_{\odot} c^{-2}}}

\newcommand\coredbentry[2]{{\tt {#1}:{#2}}}

\def\thc{{\tt THC}}
\def\snec{{\tt SNEC}}

\def\knecnn{{\tt kNECNN} } \def\skynet{{\tt SkyNet} }

\newcommand{\ie}{\textit{i.e.}}
\newcommand{\eg}{\textit{e.g.}}
\newcommand{\cf}{\textit{cf.}}

\usepackage{color}
\definecolor{cyan}{rgb}{0,0.9,0.9}
\definecolor{orange}{rgb}{0.9,0.5,0}
\definecolor{magenta}{rgb}{1,0,1}
\definecolor{purple}{rgb}{0.8,0.4,0.8}
\definecolor{gray}{rgb}{0.8242,0.8242,0.8242}

\title[]{Long-lived neutron-star remnants from asymmetric binary neutron star mergers: element formation, kilonova signals and gravitational waves}

\author[Bernuzzi {\it et al.}]{
  Sebastiano Bernuzzi$^1$,
  Fabio Magistrelli$^1$,
  Maximilian Jacobi$^1$,\and
  Domenico Logoteta$^{2,3}$,
  Albino Perego$^{4,5}$,
  David Radice$^{6,7}$
  \\
  $^1$~Theoretisch-Physikalisches Institut,
  Friedrich-Schiller-Universit{\"a}t Jena, 07743, Jena, Germany\\
  $^2$~Dipartimento di Fisica, Universit\`{a} di Pisa, Largo B. Pontecorvo, 3 I-56127 Pisa, Italy\\
  $^3$~INFN, Sezione di Pisa, Largo B. Pontecorvo, 3 I-56127 Pisa, Italy,\\
  $^4$~Dipartimento di Fisica, Universit\'{a} di Trento, Via Sommarive 14, 38123 Trento, Italy\\
  $^5$~INFN-TIFPA,Trento Institute for Fundamental Physics and Applications, via Sommarive 14, I-38123 Trento, Italy\\
  $^6$~Institute for Gravitation and the Cosmos, The Pennsylvania State
  University, University Park, PA 16802, USA\\
  $^7$Department of Physics, The Pennsylvania State University,
  University Park, PA 16802, USA\\
  $^8$Department of Astronomy and Astrophysics, The Pennsylvania State University, University Park, PA 16802, USA\\
}

\date{\today}

\pubyear{2024}

\begin{document}
\label{firstpage}
\pagerange{\pageref{firstpage}--\pageref{lastpage}}
\maketitle

\begin{abstract}
    We present 3D general-relativistic neutrino-radiation hydrodynamics
  simulations of two asymmetric binary neutron star mergers producing
  long-lived neutron stars remnants and spanning a fraction of their
  cooling time scale. The mergers are characterized by significant
  tidal disruption with neutron rich material forming a massive disc
  around the remnant. The latter develops one-armed dynamics that is
  imprinted in the emitted kilo-Hertz gravitational waves. Angular
  momentum transport to the disc is initially driven by spiral-density
  waves and enhanced by turbulent viscosity and neutrino heating on
  longer timescales. The mass outflows are composed by neutron-rich
  dynamical ejecta of mass ${\sim}10^{-3}-10^{-2}\Mo$ followed by a
  persistent spiral-wave/neutrino-driven wind of ${\gtrsim}10^{-2}\Mo$
  with material spanning a wide range of electron fractions,
  ${\sim}0.1-0.55$. Dynamical ejecta (winds) have fast velocity tails
  up to ${\sim}0.8$ (${\sim}0.4$)~c.
  The outflows are further evolved
  to days timescale using 2D ray-by-ray radiation-hydrodynamics
  simulations that include an online nuclear network. We find complete
  $r$-process yields and identify the production of $^{56}$Ni and the
  subsequent decay chain to $^{56}$Co and $^{56}$Fe. Synthetic
  kilonova light curves predict an extended (near-) infrared peak a
  few days postmerger originating from $r$-process in the
  neutron-rich/high-opacity ejecta and UV/optical peaks at a few hours
  (ten minutes) postmerger originating from weak $r$-process
  (free-neutron decay) in the faster ejecta components. Additionally,
  the fast tail of tidal origin generates kilonova afterglows
  potentially detectable in radio and X band on a few to ten years
  time scale. Quantitative effects originating from the tidal
  disruption merger dynamics are reflected in the multimessenger
  emissions. 
\end{abstract}




\section{Introduction}
\label{sec:intro}

Multimessenger observations of binary neutron star mergers (BNSMs) can help clarify the origin of heavy elements in the Universe by using coincident detections of gravitational waves and kilonovae. The latter electromagnetic transient is generated by the thermalization of radioactive decay products in the neutron-rich mass outflows~\citep{1974ApJ...192L.145L,Eichler:1989ve}. Kilonova (kN) light curves and spectra are strongly dependent on the possible NS masses and the still uncertain equation of state (EOS) of nuclear matter, \eg~\citep{Metzger:2010sy,Radice:2018pdn,Nedora:2019jhl,Jacobi:2023olu,Ricigliano:2024lwf,Fujimoto:2024cyv} and \citep{Metzger:2019zeh,Radice:2020ddv,Bernuzzi:2020tgt} for reviews. The interpretation of the diverse kN signals thus relies on robust predictions from general-relativistic simulations.

The possible NS mass range is ${\sim}1-3\Msun$~\eg~\citep{Rawls:2011jw,Ozel:2012ax}.
The upper bound can be inferred from theoretical arguments~\citep{Buchdahl:1959zz,Rhoades:1974fn,Godzieba:2020tjn} and is compatible with pulsars constraints on the ``minimum-maximum'' Tolmann-Oppenheimer-Volkhoff (TOV) mass~\citep{Demorest:2010bx,Antoniadis:2013pzd,Cromartie:2019kug} and gravitational waves (GWs) constraints, \eg~\citep{Godzieba:2020tjn}. For NS in the binaries, pulsar observations indicate mass ratios~\footnote{Here defined as the ratio between the  primary and the secondary NS mass, \ie~$q \geq 1$.}
$q\lesssim1.4$ \citep{Lattimer:2012nd,Kiziltan:2013oja,Swiggum:2015yra,Martinez:2015mya,Ferdman:2020huz}. Current GW observations are restricted to two events \citep{TheLIGOScientific:2017qsa,Abbott:2018wiz,LIGOScientific:2018mvr,Abbott:2020uma}, but indicate the possibility of even larger mass ratios. The source of GW170817 has a total mass of $M\simeq2.73-2.77\Msun$ and a mass ratio $q\lesssim1.37$ ($q\sim1.89$) for low (high) spin priors \citep{Abbott:2018wiz}. The source of GW190425 is associated with the heaviest BNS source known to date with $M\simeq 3.2-3.7\Msun$ and the mass ratio can be as high as $q\lesssim1.25$ ($q\lesssim2.5$) \citep{Abbott:2020uma}.

The observational signatures of BNSMs depend critically on the NS masses and the mass asymmetry. In particular, the tidal disruption of the secondary NS in asymmetric BNSM can significantly affect the merger remnants~\citep{Rosswog:2000nj,Shibata:2003ga,Shibata:2006nm,Kiuchi:2009jt,Rezzolla:2010fd,Dietrich:2015pxa,Dietrich:2016hky,Lehner:2016lxy,Bernuzzi:2020txg,Perego:2021mkd}. Prompt black hole formation for mass ratio $q\gtrsim1.4$ happens at a threshold mass lower than the equal-mass case~\footnote{See \eg~\citep{Hotokezaka:2011dh,Bauswein:2013jpa,Kashyap:2021wzs} for equal-mass prompt collapse studies.} as a result of the accretion of material from the secondary to the primary star. This behaviour is strongly dependent on the NS equation of state (EOS); it cannot be parameterized in a simple way using basic properties of TOV solution because it depends, for example, on the EOS incompressibility at the maximum TOV density \citep{Perego:2021mkd}. The tidal disruption of the secondary star also produces accretion discs with baryon masses ${\sim}0.15\Msun$ \citep{Bernuzzi:2020txg}. The latter is significantly heavier than remnant discs in equal-masses prompt collapse mergers or even short-lived remnants, \eg~\citep{Radice:2018pdn,Nedora:2019jhl}. Simulations of asymmetric BNSM suggest that these merger events are associated with bright and temporally extended kNe. The latter are expected to be particularly luminous in the red and (near)-infrared electromagnetic bands due to the nucleosynthesis of lanthanides in the neutron-rich tidal ejecta \citep{Rosswog:2017sdn,Wollaeger:2017ahm,Bernuzzi:2020txg}.

Asymmetric mergers with mass ratios up to $q\gtrsim1.4$ and sufficiently stiff EOS can also produce NS remnants that are (at least temporarily) stable against gravitational collapse, although this scenario has been less investigated so far. Similarly to the equal mass case, the presence of a NS remnant can alter the remnant disc properties, both in terms of compactness and composition, \eg~\citep{Perego:2019adq,Camilletti:2024otr}. Consequently, also the kN emission is influenced. Angular momentum transport from the remnant's spiral-arms powers a viscous ejecta component about the orbital plane that increases the kN brightness~\citep{Radice:2018pdn,Nedora:2019jhl}. Neutrino fluxes from both the remnant and the disc power strong neutrino-driven winds towards the polar regions \citep{Dessart:2008zd,Perego:2014fma,Metzger:2014ila,Martin:2015hxa,Fujibayashi:2020qda}. The nucleosynthesis in this more proton-rich matter produces $r$-process elements with mass number $A<130$ \citep{Martin:2015hxa,Just:2023wtj} as well as light elements \citep[\eg][]{Perego:2020evn,Chiesa:2023jno}. The associated kN emission peaks in the ultraviolet/optical bands at timescales of hours-days.

In this work, we present new results from asymmetric BNSM using ab-initio $(3+1)$D numerical-relativity simulations. We focus on two binaries configurations with large mass asymmetry that do not promptly collapse but form a NS remnant. The merger dynamics is characterized by a significant tidal disruption of the secondary star. Our simulation package, described in Sec.~\ref{sec:meth}, includes state-of-art microphysical EOS, a gray truncated momentum scheme for neutrino transport and a subgrid model for magnetically-induced turbulence. The  $(3+1)$D ejecta data are further evolved to days timescales with a ray-by-ray radiation-hydrodynamics code that includes an online nuclear network~\citep{Magistrelli:2024zmk}.

The paper is structured as follows.
In Sec.~\ref{sec:meth} we present the simulation methods.
In Sec.~\ref{sec:dyn} we discuss the remnant and ejecta dynamics up to hundreds of milliseconds postmerger. 
In Sec.~\ref{sec:nucleo}, we discuss element production in the outflows. 
In Sec.~\ref{sec:obs}, we discuss the long-term outflow evolution and the kN light curves, together with the gravitational waves emission.
Conclusions follows.

CGS units are employed everywhere except for masses (reported in solar masses, $\Msun$), lengths (in km), temperature (in MeV), entropy per baryon (in units of the Boltzmann constant per baryon and indicated as $k_{\rm B}$).
Geometric units $c=G=1$ are employed for the GW strain.

\section{Simulation methods}
\label{sec:meth}

We present results for two asymmetric binaries with mass ratio $q\simeq1.77$ and $q\simeq1.49$ and total baryon mass of, respectively, $M_b=3.190\Mo$ and $M_b=3.064\Mo$. 
Matter is described by the microphysical EOS DD2~\citep{Typel:2009sy,Hempel:2009mc} and BLh \citep{Bombaci:2018ksa,Logoteta:2020yxf}. These EOS include neutrons, protons, nuclei, electrons, positrons, and photons as relevant thermodynamics degrees of freedom. The EOS models have nuclear matter parameters at saturation density broadly compatible with observational and experimental constraints. Cold, neutrino-less $\beta$-equilibrated matter described by these microphysical EOS predicts NS maximum masses and radii within the range allowed by current astrophysical constraints.
The gravitational masses of the simulated binaries are $M\simeq(1.80+1.08)\Mo$ and $M\simeq(1.635+1.146)\Mo$ for the DD2 and BLh EOS respectively.

Constraint-satisfying initial data for the simulations are produced assuming an irrotational binary in quasi-circular orbit (\ie~imposing a helical Killing vector), conformally flat metric and matter in neutrino-less beta- and hydrostationary equilibrium. They are generated using the publicly available pseudo-spectral multi-domain library \texttt{Lorene} \citep{Gourgoulhon:2000nn}.

The evolution of the initial data are performed using $(3+1)$D general-relativistic radiation-hydrodynamics simulations up to $\sim100$~ms postmerger. The spacetime evolution employs the Z4c free-evolution scheme for Einstein's equations \citep{Bernuzzi:2009ex,Hilditch:2012fp}. The gauge sector is solved using the ``1+log'' equation for the lapse and the Gamma-driver equation for the shift. The general-relativistic hydrodynamics (GRHD) equations are formulated in conservative form (see~\citet{Radice:2018pdn} for details on the precise equations solved here) and augmented by a large-eddy-scheme that accounts for angular momentum transport due to magnetohydrodynamics effects \citep{Radice:2017zta, Radice:2020ids, Radice:2023zlw}. Turbulent viscosity is parametrized in terms of a characteristic density-dependent length scale $\ell_{\rm mix}(\rho)$ that is modeled after the high-resolution BNS data of \citet{Kiuchi:2017zzg,Kiuchi:2022nin}. We simulate with two $\ell_{\rm mix}$ models, named K1~\citep{Radice:2020ids} and K2~\citep{Radice:2023zlw}, where K2 extends to densities as low as $\rho\sim10^8\gccm$, see \citet{Radice:2023zlw} for details. We remark that the K2 models, being computed from high-resolution, long-term, and ab-initio GRMHD simulations of mergers with initial magnetic field intensities of $\sim10^{15}$~G, provide a realistic upper limit for the viscosity inside the remnant and the disc. Neutrino radiation is simulated with a truncated, two-moment gray scheme that retains all nonlinear neutrino-matter coupling term~\citep{Radice:2021jtw}. The scheme employs the Minerbo closure and considers three different neutrino species: electron neutrinos $\nu_e$, anti-electron neutrinos $\bar{\nu}_e$, and a collective species $\nu_x$ describing heavy flavour neutrinos and antineutrinos. The set of weak reactions is the same as described and tabulated in our previous work, see \eg~\citet{Galeazzi:2013mia,Radice:2016dwd,Perego:2019adq}.

The $(3+1)$D simulations are performed with the \thc~code~\citep{Radice:2012cu,Radice:2013hxh,Radice:2013xpa,Radice:2015nva,Radice:2016dwd,Radice:2021jtw}, which is built on top of the \texttt{Cactus} framework \citep{Goodale:2002a,Schnetter:2007rb}. The spacetime is evolved with the \texttt{CTGamma} code \citep{Reisswig:2013sqa} which is part of the \texttt{Einstein Toolkit} \citep{Loffler:2011ay}. The time evolution is performed with the method of lines, using fourth-order finite-differencing spatial derivatives for the metric and the strongly-stability preserving third-order Runge-Kutta scheme \citep{Gottlieb:2009a} as the time integrator. The timestep is set according to the Courant-Friedrich-Lewy criterion with a factor $0.15$. Berger-Oliger conservative adaptive mesh refinement \citep{Berger:1984zza} with sub-cycling in time and refluxing is employed \citep{1989JCoPh..82...64B,Reisswig:2012nc}, as provided by the \texttt{Carpet} module of the \texttt{Einstein Toolkit} \citep{Schnetter:2003rb}.

The simulation domain is a cube of side ${\sim}3024$ km, centred at the centre of mass of the binary system. Only the $z\geq0$ portion of the domain is simulated and reflection symmetry about the $xy$-plane is used for $z<0$. The grid consists of 7 refinement levels centred on the two NSs or in the merger remnant, with the finest level covering entirely each star. Simulations are performed at three resolutions that are identified by the grid spacing of the finest refinement level at the start of the simulation. Following the notation of our previous papers, we use ${\Delta x}_\mathrm{VLR} \approx 494$~m (VLR), ${\Delta x}_\mathrm{LR} \approx 247$~m (LR), ${\Delta x}_\mathrm{SR} \approx 185$~m (SR). Simulations at VLR are typically insufficient to obtain quantitative remnant properties/evolutions \citep{Zappa:2022rpd}; they were thus conducted for shorter evolution times and are not further discussed here.

Mass ejecta profiles extracted from the $(3+1)$D simulations are further evolved to days timescales using a ray-by-ray Lagrangian radiation-hydrodynamics approach that includes online nuclear network~\citep{Magistrelli:2024zmk}. The initial Lagrangian profiles are constructed from the 3D ejecta using the procedure described in \cite{Wu:2021ibi}. These profiles are evolved with the \snec~code \citep{Morozova:2015bla,Wu:2021ibi} coupled to the \skynet nuclear network (NN) \citep{Lippuner:2017tyn}, as described in \citet{Magistrelli:2024zmk}~(\knecnn~hereafter.) The NN includes 7836 isotopes up to~$^{337}$Cn and uses the JINA REACLIB \citep{Cyburt:2010a} and the same setup as in \cite{Lippuner:2015gwa, Perego:2020evn}. Hence, our simulations incorporate self-consistently the heating due to the nuclear burning and hydrodynamics couplings.

The nuclear energy thermalization scheme of \knecnn~includes contributions from $\gamma$ rays, $\alpha$ particles, electrons, and other nuclear reactions products. The thermalization factor for the $\gamma$'s is calculated as in \cite{Hotokezaka:2019uwo, Combi:2022nhg}, starting from the detailed composition of the ejecta and the same effective opacity tables from the \href{https://www.nist.gov/pml/xcom-photon-cross-sections-database}{NIST-XCOM} catalogue \citep{Berger:2010} used in \citet{Barnes:2016umi}. The opacity data are combined with energy spectra for the emission for $\gamma$-rays, which is reported for each specific isotope in the \href{https://www-nds.iaea.org/exfor/endf.htm}{ENDF/B-VIII.0} database \citep{Brown:2018jhj}. We use this information to distribute the nuclear power coming from the NN into the different decay products (see \cite{Magistrelli:2024zmk} for further details).
For electrons and $\alpha$ particles the analytic expressions of \citet{Kasen:2018drm} are used.
Electrons and positrons are thermalized in the same way.
The nuclear energy emitted in neutrinos is explicitly removed from the system. This energy is associated with the neutrinos produced by the NN. Note however that the information about the neutrino fluxes coming from the remnant is not transferred to the network; its cumulative effect is taken into account in the 3D simulations. Neutrinos in regions with radii ${\gtrsim}400$km are free streaming~\citep{Endrizzi:2019trv} and their contribution is expected to be a comparably small correction to the final integrated abundances and heating rates~\citep{Goriely:2015fqa}.
  
The rest of the nuclear power, associated \eg~with fission fragments, daughter nuclei and $X$-rays, is thermalized with an effective thermalization factor of 80\%. %
To calculate the luminosity, we employ the same analytic, time-independent and $Y_e$-dependent gray opacity introduced in \citet{Wu:2021ibi}.

The angular dependency of the ejecta properties is approximately included in a ray-by-ray fashion. We use $51$ angular bins in the polar angle $\theta\in(0,\pi)$, where $\theta=0$ is the pole above the remnant and $\theta = 90^{\circ}$ identifies the orbital plane, and average the ejecta properties along the azimuthal angle $\phi\in(0,2\pi)$.
For each angular bin, the initial profile is mapped into an effective 1D problem by multiplying the total mass by the scaling factor $\lambda_\theta = 4\pi / \Delta\Omega$, where $\Delta\Omega \simeq 2\pi \sin{\theta} \, d\theta$ represents the solid angle included in the angular bin. This ensures that all the intensive quantities (including density) retain their original values. The different angular sections are then independently evolved by discretizing the 1D radiation-hydrodynamic equations over $n = 1000$ fluid elements. Non-radial flow of matter and radiation are neglected. The 1D results for each bin are recombined by keeping the intensive quantities unchanged and rescaling the extensive ones by $1/\lambda_\theta$. In particular, the global mass fractions and abundances are calculated with a mass-weighted average over all the mass shells and angular bins. The kilonova light curves are recombined accounting for the viewing angle as in \citet{Martin:2015hxa,Perego:2017wtu}.

\section{Remnant and Mass outflows}
\label{sec:dyn}

\subsection{Merger Remnant}
\label{sec:rem}

\begin{figure}
  \centering
  \includegraphics[width=0.49\textwidth]{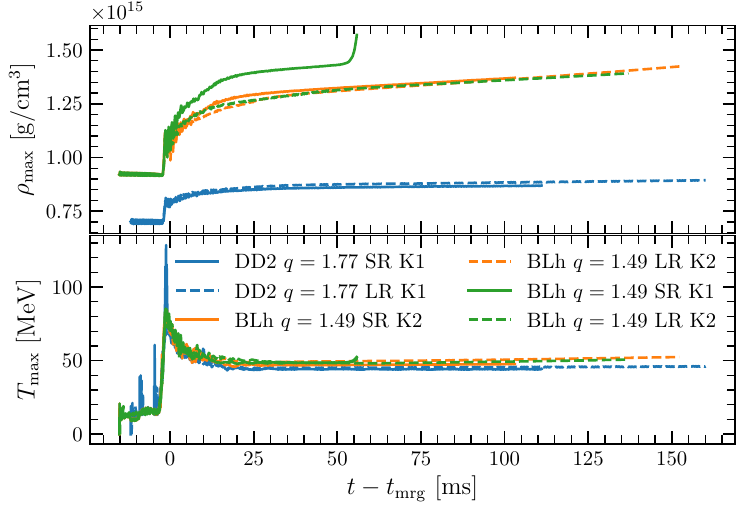}
  \caption{Evolution of the maximum temperature and maximum rest-mass density for all the simulations at grid resolutions LR and SR.
    The simulation BLh $q=1.49$ with K1 turbulent viscosity at
    resolution SR collapse at $t-t_{\rm mrg}\sim50$~ms, see text for
    details.}
  \label{fig:rho_temp_max}
\end{figure}

Both binaries undergo a tidally disruptive merger during which the secondary star accretes material on the primary star. The remnants do not prompt collapse to a black hole although the BLh binary is close to the prompt collapse mass threshold of $M\simeq2.9\Mo$ having a binary mass of $M\simeq2.78\Mo$ \citep{Perego:2021mkd}. The merger products are long-lived neutron stars stable on timescales of ${\gtrsim}100$~ms, the BLh binary eventually collapses at ${\sim}114$~ms postmerger. They are surrounded by an envelope (``disc'') that is sustained by angular momentum transport and neutrino heating from the central object. Gravity and neutrino cooling drive instead disc accretion and the whole remnant towards a more compact state. These dynamics create the conditions for the development of massive winds from the disc that are flued by angular momentum transport from the central object and neutrino irradiation.

Figure~\ref{fig:rho_temp_max} shows the evolution of the maximum rest-mass density and temperature for all the simulations. The maximum density raises at merger as a consequence of the mass accretion from the secondary star. Due to the softer EOS and despite the smaller mass, the BLh binary is more compact and its remnant reaches maximum densities about 60\% higher than the DD2. The maximum temperatures spike to ${\sim}80-140$~MeV and are generated at densities $\rho\sim10^{14}\gccm$ from a shock developing between the accreting matter and the core of the primary star. These shocks activate weak processes and produce charateristic neutrino luminosity peaks of ${\gtrsim}10^{52}$~erg/s. The neutrino luminosities and average energies are similar to the equal-mass case, \eg~\citet{Zappa:2022rpd}.
After the spike, the maximum temperatures in the remnant remain of the order of ${\sim}50$~MeV during the entire postmerger evolution, being continuously fed by shock waves from spiral density waves (more below). As visible in the plot, the BLh binary simulated at resolution SR and with the K1 turbulent viscosity scheme collapses at $t-t_{\rm mrg}\sim50$~ms. While all the BLh remnants may collapse anytime on these timescales, higher resolutions would be needed to confirm the K1 results, \cf~\citet{Zappa:2022rpd}. In the following, we focus on the BLh simulation with the K2 turbulent viscosity scheme which gives consistent results among the resolutions. With the K2 scheme, angular momentum redistribution is more efficient than for the K1 scheme, and thus it favors the stability of the remnant against collapse. In our discussion, we conventionally define the NS remnant as the central object at rest-mass densities $\rho>10^{13}\gccm$ and refer to the remaining bound material as to the disc. We note that at the NS remnant densities the neutrino mean free path becomes smaller than any relevant length scale over which termodynamics quantities change significantly. Thus, neutrinos form a trapped gas in equilibrium with the fluid and diffuse out on the diffusion timescale \citep{Perego:2014fma,Foucart:2015gaa,Endrizzi:2019trv,Espino:2023dei}.

\begin{figure*}
  \centering
  \includegraphics[width=0.99\textwidth]{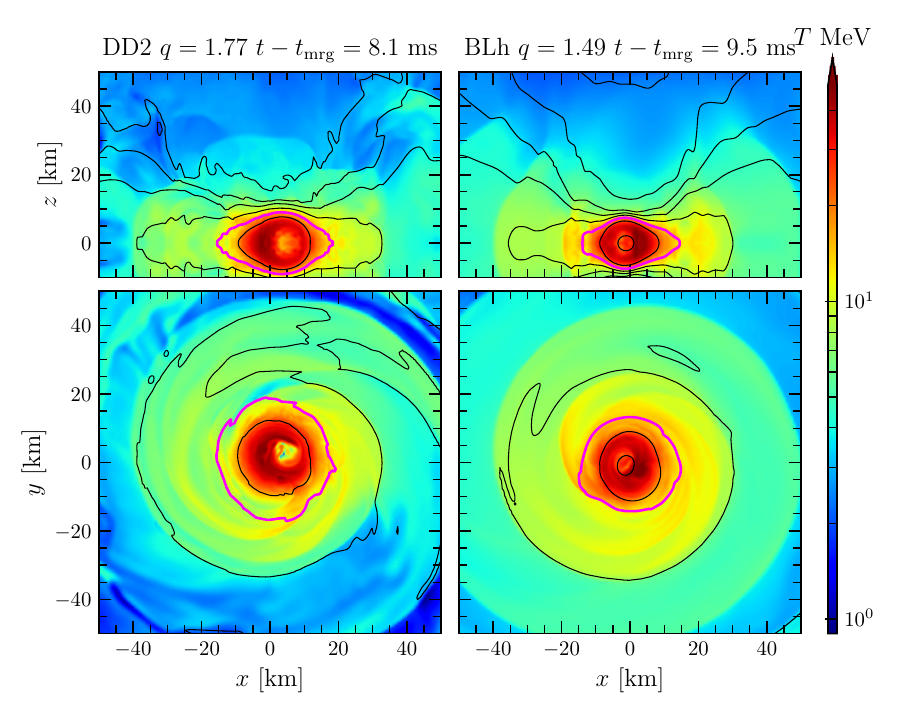}
  \caption{Remnants' 2D snapshots of the temperature (color-coded) and the rest-mass density (isocontours) for simulations at SR resolution.
    The magenta contour refers to rest-mass densities of $10^{13}\gccm$, the black contours refer to $10^{15},10^{14},10^{12},...,10^{8}\gccm$.}
  \label{fig:remnants2d}
\end{figure*}

The rotation of the NS remnant produces density spiral arms that transport angular momentum outwards into the disc~\citep{Nedora:2019jhl}. This process is enhanced by turbulent viscosity and generates density waves propagating into the disc. Figure~\ref{fig:remnants2d} shows temperature snapshots in the orbital plane $(x,y)$ and above the remnant, plane $(x,z)$. Spiral density waves develop just outside the NS remnant (magenta contours) in the $(x,y)$ plane and correspond to local maxima of temperature (outside the magenta contours). The shock waves are also well identifiable in the $(x,z)$ plane, right-top panel. As noted above, the figure shows that the maximum temperatures are reached within the remnant but not at the maximum densities, see \cite{Perego:2019adq} for a detailed analysis of the equal-mass case. The DD2 NS remnant has a colder core, from the matter of the primary star. The BLh remnant develops stronger shock waves than the DD2 remnant as a result of less disruptive dynamics of the secondary star and a more violent merger of the two NS cores. However, the BLh spiral waves transfer more energy and shut off more quickly with time. Contrary, the DD2 spiral waves appear weaker but more persistent. The spiral wave dynamics has a strongly non-axisymmetric one-armed component \citep{Paschalidis:2015mla,East:2015vix,Radice:2016gym} appearing as ``left-right'' oscillations of the NS remnant in the $(x,z)$ plane. This $m=1$ mode dominates and persists over the entire simulation, while the $m=2$ bar-mode component is efficiently damped by gravitational radiation. We will return to this point in Sec.~\ref{sec:gws}, in connection to the gravitational-wave emission.

\begin{figure*}
  \centering
  \includegraphics[width=0.99\textwidth]{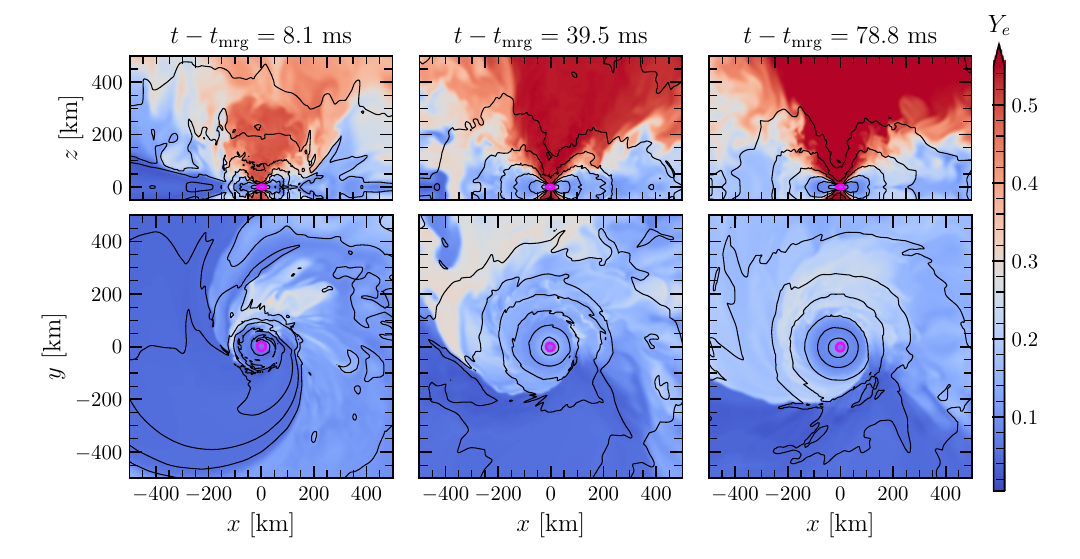}
  \caption{2D snapshots of the electron fraction (color-coded) and rest-mass density (isocontours) at different postmerger times for the DD2 $q=1.77$ SR simulation. The magenta contour refers to rest-mass densities of $10^{13}\gccm$, the black contours refer to $10^{12},10^{11},10^{10},...,10^{5}\gccm$.}
  \label{fig:ye2d:DD2}
\end{figure*}

\begin{figure*}
  \centering
  \includegraphics[width=0.99\textwidth]{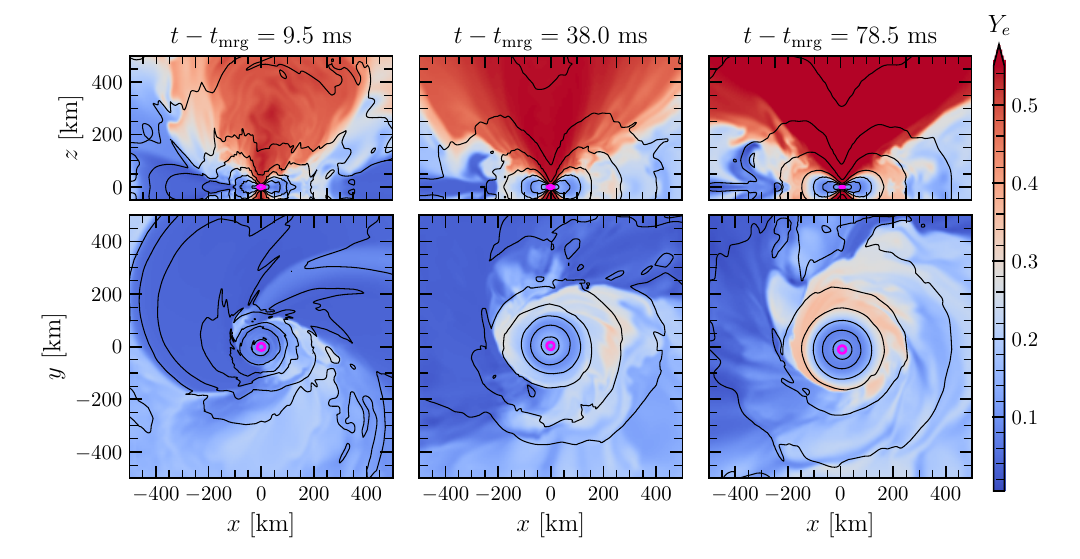}
  \caption{Same as Fig.~\ref{fig:ye2d:DD2} but for BLh $q=1.49$ SR K2 simulation.}
  \label{fig:ye2d:BLh}
\end{figure*}

The disc is not only fed with angular momentum by the spiral density waves, but it is also irradiated by neutrinos from the NS remnant and from the inner layers of the disc itself. Figure~\ref{fig:ye2d:DD2} shows the electron fraction of the disc on the orbital and $(x,z)$ planes at three different postmerger times for the DD2 simulation. Figure~\ref{fig:ye2d:BLh} is the same for the BLh simulation. Neutrinos streaming from their diffusion spheres, which are reasonably approximated by rest-mass density isocontours of $\rho_{\nu\ \rm diff}\gtrsim10^{11}\gccm$ \citep{Perego:2014fma,Endrizzi:2019trv}, protonize the disc's matter. At densities $\rho\lesssim10^{10}\gccm$, corresponding to distances of ${\sim}200$~km on the orbital plane, the electron fraction reaches typical values $Y_e\sim0.3$ in the DD2 simulations. Above the remnant, neutrino absorption is maximal and a funnel with polar angles $\theta>45^{\circ}$ develops shortly after the merger. Here, the electron fraction reaches equilibrium values $Y_e\gtrsim0.42$ but baryon densities are lower than in the equatorial regions, $\rho\lesssim10^7\gccm$. These processes are very similar for the two binary's remnants, but the BLh remnant develops a higher-$Y_e$ region around the orbital plane (radii ${\sim}200$~km.)

The disc expansion eventually leads to the evaporation of the outer layers of the disc. Disc winds start at equatorial radii of a few hundreds of kilometers. There, the material becomes unbound due to the
$\alpha$ (iron-group) recombination and the subsequent liberation of about $8$ ($7$)~MeV/nucleon of nuclear binding energy \citep{Beloborodov:2008nx,Metzger:2008av,Lee:2009uc,Just:2023wtj}. The evaporation in our simulations is aided and sustained by neutrino captures, especially above the remnant (polar angles $\theta\gtrsim60^{\circ}$) where the baryon densities are progressively lower. The disc reaches a turbulent state in these regions, roughly corresponding to rest-mass density isocontours of $\rho\sim10^{9}\gccm$.

\subsection{Mass outflows}
\label{sec:ejecta}

The tidal disruption and the remnant's viscous dynamics described above power massive outflows of $M_{\rm ej}\sim10^{-2}\Mo$ over a timescale of ${\sim}100$~ms. The outflows are characterized by a neutron-rich dynamical component and a wind spanning a wide range of proton fractions. Outflows are calculated during the simulation at a coordinate sphere centered in the origin of the grid and with radius $r=740$~km. Fluid elements are flagged as unbound if their velocity is directed outward pointing normal of the sphere and $h u_t \leq -1$ (Bernoulli criterion), where $h$ is the fluid-specific enthalpy and $u^t$ the 0-component of the fluid's 4-velocity. The properties of such ejecta are described in the following.

\begin{figure*}
  \centering
  \includegraphics[width=0.99\textwidth]{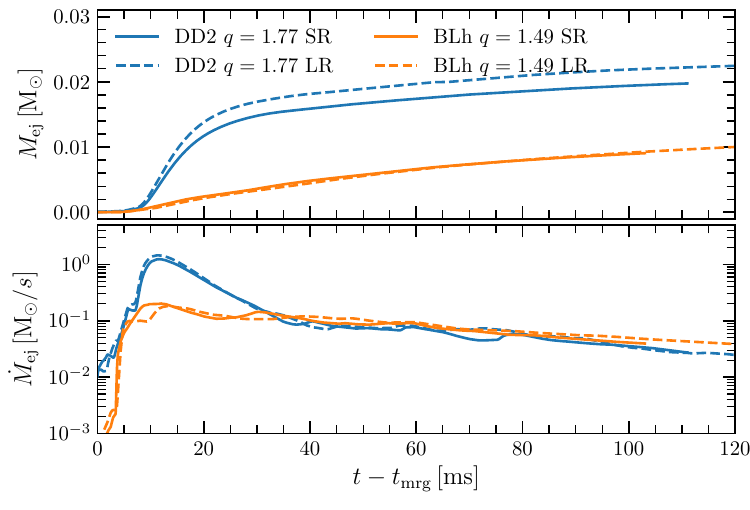}
  \caption{Evolution of the outflows' mass (top) and outflows' rate (bottom) for the simulations at grid resolutions LR and SR. The outflowed mass rate peaks at early times are due to tidal mass ejection during merger. Afterwards, the outflow is dominated due to spiral-wave and neutrino-driven winds. At $t-t_{\rm mrg}\sim40$ ms the outflow rate is $M_{\rm ej}\sim0.1 \Mo/s$ and it decreases to ${\sim}0.02\Mo/s$ at later times as ${\sim}1/t$.}
  \label{fig:masses}
\end{figure*}

Figure~\ref{fig:masses} shows the ejecta mass evolutions for the two binaries. The outflowed mass rate peaks at early times are due to the tidal mass ejection at merger timescales. The peak is more prominent for the DD2 binary because of the almost complete tidal disruption of the secondary star. Afterwards, the outflow is powered by the density spiral waves around the orbital plane and by neutrino winds developing above the remnant. 
The winds persist during the entire simulated time. They power mass outflows at a rate of $\dot{M}_{\rm ej}\sim0.1 \Mo$/s at $t-t_{\rm rmg}\simeq40$~ms, which decreases to ${\sim}0.02\Mo/s$ as $\dot{M}_{\rm ej}(t){\sim}1/t$. The mass outflow from the neutrino wind is quantitatively very similar in the DD2 and BLh binaries, despite the significant differences in mass ratio and EOS.

Higher grid resolutions decrease the outflowed mass because a better-resolved flow typically leads to more compact remnants \citep{Zappa:2022rpd}. The differences between resolutions SR and LR are of order ${\sim}10$\%, which are comparable to the differences due to the finite extraction sphere, \eg~at radii $600$~km and $740$~km. Assuming first-order convergence with grid resolution, the computed mass for SR appears very robust for the simulated physics processes. Magnetic-field instabilities and magnetic pressure are not fully accounted for here, but they can enhance the mass of SR runs up to a factor two-to-ten on the simulated timescales and for extreme (magnetar-strength) intensities, \eg~\citep{Mosta:2020hlh,Kawaguchi:2022bub,Combi:2022nhg}. Nuclear heating following $r$-process nucleosynthesis could further increase the ejecta mass on second timescales due to the deposition of a few MeV/nucleon in the fluid \citep{Rosswog:2013kqa,Foucart:2021ikp}. Overall, the mass computed here is a solid lower bound for the complete ejecta mass emitted on longer timescales.

\begin{figure*}
  \centering
  \includegraphics[width=0.49\textwidth]{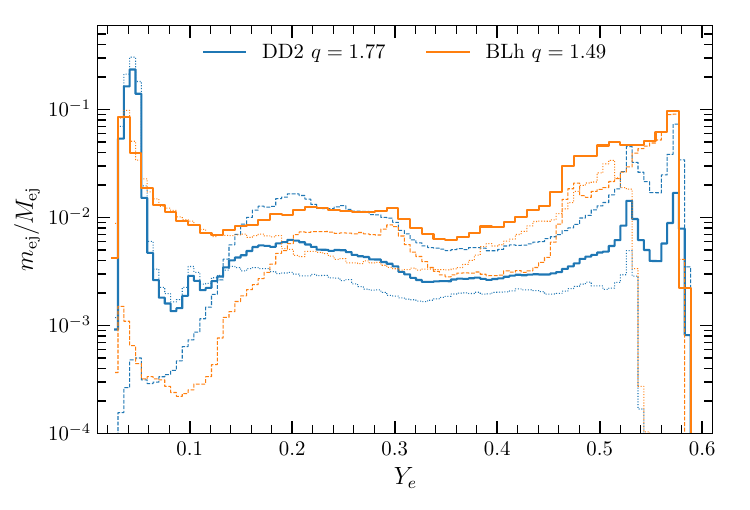}
  \includegraphics[width=0.49\textwidth]{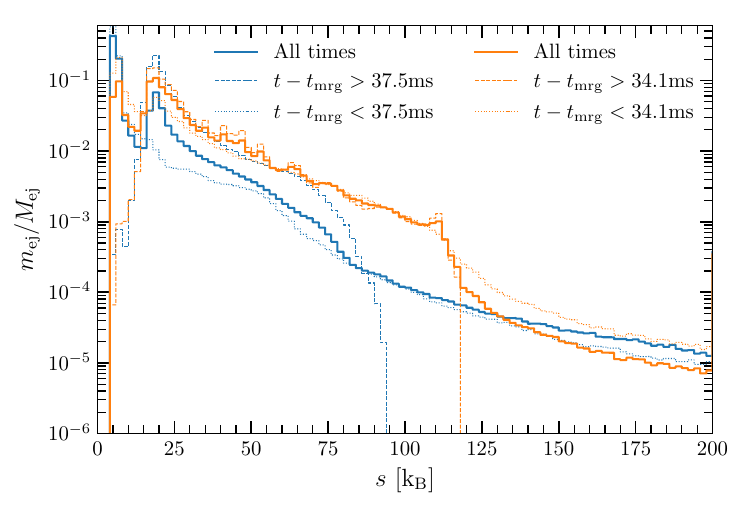}\\
  \includegraphics[width=0.49\textwidth]{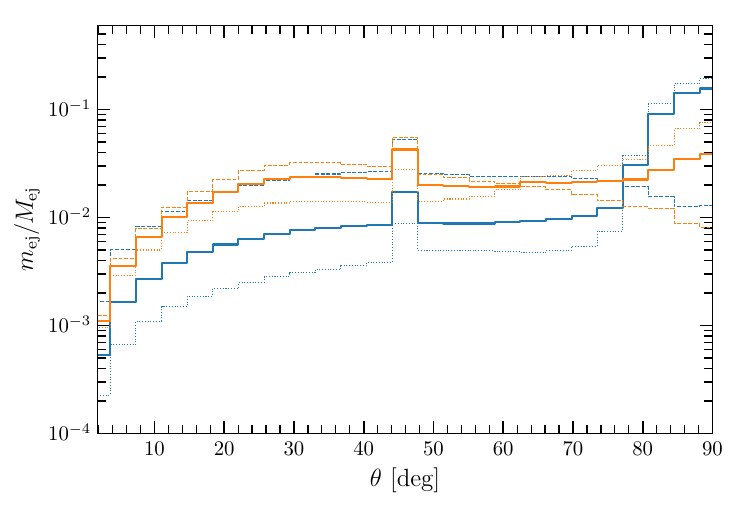}
  \includegraphics[width=0.49\textwidth]{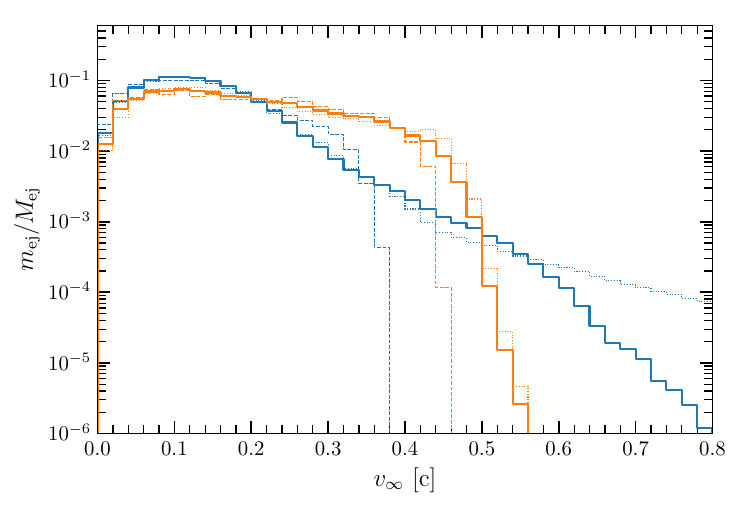}
  \caption{Mass-weighted histograms of the properties of the outflows for the simulations at grid resolutions SR. The different panels show the electron fraction $Y_e$, the asymptotic velocity $v_{\infty}$, the polar angle $\theta$, the entropy per baryon $s$ in units of the Boltzmann constant. Results from three analyses are shown: at all times, until times shortly after the tidal ejecta, and from times after the tidal ejecta (respectively solid, dashed and dotted lines).
    The outflows' fluid elements are characterized by very neutron-rich material due to the tidal ejection and an approximately uniform distribution extending to $Y_e\gtrsim0.5$ due to the subsequent remnants dynamics. The BLh $q=1.49$ binary produces outflows with a very proton-rich content due to contributions from both the core shock and the neutrino-driven wind.
    The DD2 $q=1.77$ binary produces outflows with fast tidal tails up to ${\sim}0.8\,c$. Tidal ejecta are generated around the equatorial plane, while later winds are generated up to higher latitudes. The peaks at $\theta=45^{\circ}$ are a binning artifact.} 
  \label{fig:outflow_hist}
\end{figure*}

The properties of the outflows are quantified in Fig.~\ref{fig:outflow_hist}, which shows the mass-weighted histograms of the electron fraction, the ejecta's asymptotic velocity, the polar angle and the entropy per baryon (see \eg~Eq.~(5) of \citet{Nedora:2020hxc} for the precise expression for these histograms).
The dynamical tidal ejecta are characterized by low proton fraction $Y_e<0.1$ and low entropy $s<10$~MeV. This material originates from the secondary star and it is expelled around the orbital plane within a polar angle of $\theta\lesssim 20^{\circ}$. Owning to the disruption dynamics, the mass distribution is not uniform in the azimuthal angle $\phi$ but the ejecta emerge from a particular direction (\cf~Fig.12 of \citet{Bernuzzi:2020txg}.) Part of the tidal ejecta is accelerated by tidal torque to high velocities $v_\infty\gtrsim0.6$, although this is dependent on the mass ratio and the tidal polarizability parameters of the secondary star. Fast tails are significantly slower for the BLh binary than for the DD2 binary. Although the material in the fast tail comes from tidal ejecta, its eletron fraction is typically higher than the slower part of the dynamical ejecta with $Y_e\gtrsim0.22$.
\citet{Bernuzzi:2020txg} found that fast tails are generically suppressed for large mass ratios $q\gtrsim1.5$ in accretion-induced prompt collapse mergers. On the one hand, the result is confirmed here for the BLh binary, which was also simulated there. On the other hand, the new simulations indicate that a binary with a stiffer EOS (like DD2, not simulated in \citet{Bernuzzi:2020txg}) combined with a sufficiently large mass ratio can also generate fast tails. 

The wind ejecta are characterized by a broad range of proton fractions extending to $Y_e\sim0.5$ and entropy per baryon peaking at $s\sim20$ k$_{\rm B}$. This material is irradiated by neutrinos emitted by the remnant with progressively more intense neutrino fluxes as the latitude increases, \ie~away from the orbital plane, see Fig.~\ref{fig:ye2d:DD2} and \cite{Perego:2014fma}. However, the baryon density reduces at small polar angles, and the wind ejecta emerge mostly up to polar angles of $\theta\gtrsim30^{\circ}$. 
The BLh $q=1.49$ binary produces outflows with a more proton-rich content than the DD2 due to contributions from both the core shock and the neutrino-driven wind. This contribution emerges from densities $10^{8}\gccm\lesssim\rho\lesssim10^9\gccm$, as clear from comparing the right column of Fig.~\ref{fig:ye2d:DD2} and Fig.~\ref{fig:ye2d:BLh}. 
The irradiation from the disc is expected to be further enhanced by neutrino pairs annihilation \citep{Dessart:2008zd,Zalamea:2010ax,Perego:2017fho}, which is not included in our simulations.
The wind ejecta have asymptotic velocities $v_\infty\lesssim0.4$~c, which decrease with evolution time. This is mostly due to the progressively weaker spiral-wave wind in the orbital plane and the slower neutrino winds above the remnant \citep{Perego:2017fho}.
\citet{Radice:2023zlw,Radice:2023xxn} have simulated a DD2 binary with the same method used here and equal-mass  $M=(1.35+1.35)\Mo$~\footnote{Note also the chirp masses differ, $\M_c(q=1.77)\simeq1.20\Mo$ and $\M_c(q=1)\simeq1.17\Mo$.}. Comparing to the DD2 $q=1.77$ data, the ejecta of that equal mass simulation has no neutron-rich tidal ejecta, the shocked-heated material at intermediate latitudes ($\theta\lesssim45^{\circ}$) has higher electron fraction $Y_e\gtrsim0.22$, the spiral wave wind is less persistent in time and spans a narrower range of electron fractions ($Y_e\gtrsim0.22$), and entropies above the remnant can be a factor two larger. The ejecta of the equal-mass DD2 $M=(1.35+1.35)\Mo$ simulation are indeed quantitatively more similar to the BLh $q=1.49$ which has a less extreme mass ration and a closer chirp mass ($\M_c(q=1.49)\simeq1.188\Mo$.) 

In summary, tidal disruption mergers producing a remnant NS have mass outflows composed of a tidal component and a component powered by the remnant's spiral waves and by neutrino heating. The wind component is persistent over the cooling timescale of the remnant and slowly decelerates as the spiral-arms dynamics weaken. The material composition spans the entire range of proton fractions.

\section{Nucleosynthesis}
\label{sec:nucleo}

\begin{figure*}
  \centering
  \includegraphics[width=\textwidth]{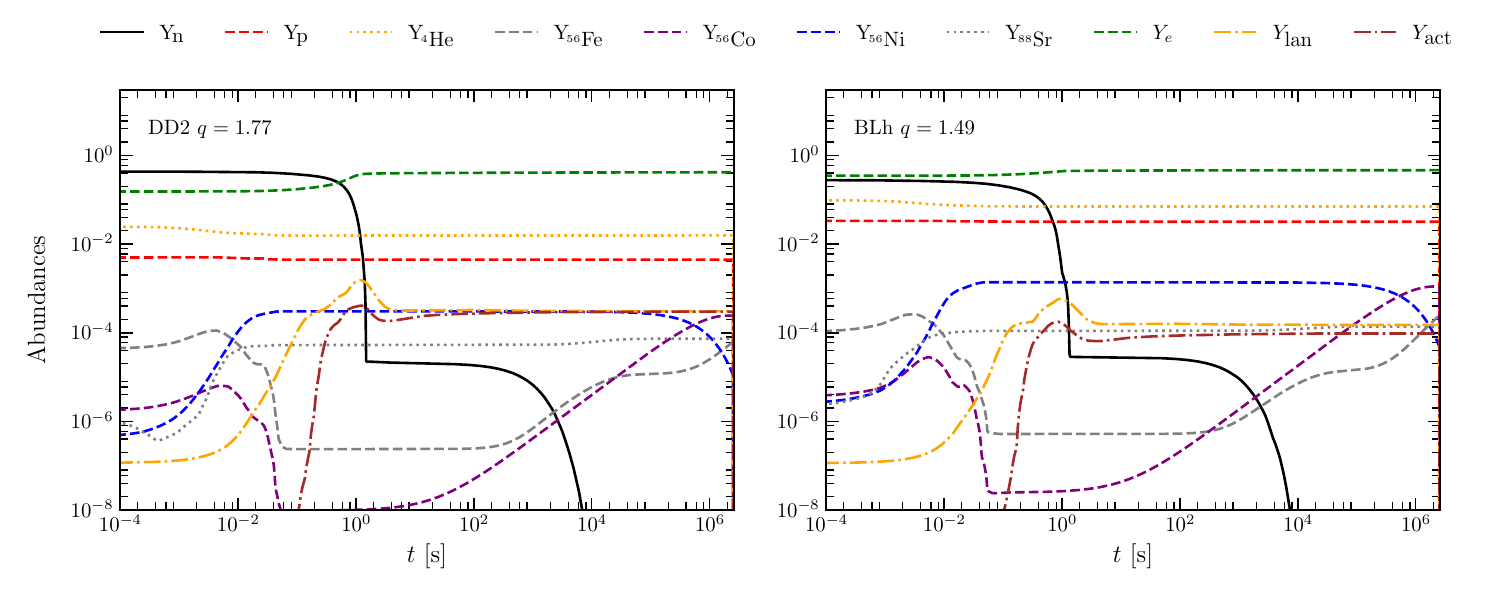}
  \caption{Evolution of the abundance of some selected elements for
    the simulations at grid resolution SR as obtained with
    \knecnn. In this and following
      plots, $t=0$ is the last simulation time at which the ejecta is
      recorded on the extraction sphere and mapped to \knecnn.}
   \label{fig:elements}
\end{figure*}

Heavy nuclei are produced in the ejecta by rapid neutron capture on timescales of ${\lesssim}1\,$s. After these times, neutron captures become inefficient and nuclei start to stabilize mostly via $\beta$-decays. The decay products thermalize, heating the fluid and feeding the kilonova (Sec.~\ref{sec:kn}). Our \knecnn~simulations capture this process self-consistently, under the assumption of axisymmetry using a 2D ray-by-ray setup~\citep{Magistrelli:2024zmk}.

Figure~\ref{fig:elements} shows the evolution of the abundances for free protons ($Y_p$), neutrons ($Y_n$) and a few selected isotopes. In particular, we show the most abundant $^4$He, the Ni-Co-Fe decay chain, ${}^{88}$Sr and the cumulative abundances of lanthanides and actinides. Abundances are calculated as mass-weighted averages over the ejecta.
The drop in $Y_n$ at around $t\sim 1$~s indicates the end of the $r$-process nucleosynthesis. The remaining free neutrons $\beta$-decay on a timescale of ${\sim}10$~minutes, causing $Y_n$ to further drop below $10^{-8}$ at $t\sim1$~hour. Initial neutron abundances are larger for the DD2 binary than for the BLh binary due to the more massive dynamical ejecta originating from tidal disruption (part of this material has also faster velocity tails).
The production of lanthanides and actinides stops at the end of the $r$-process nucleosynthesis, while $^{56}$Ni, $^{88}$Sr are already fully produced at $t\sim 10^{-2}$~s. During the early phases of nucleosynthesis, $^{56}$Co and $^{56}$Fe are first produced at nuclear statistical equilibrium freeze-out and then consumed as seed nuclei. They start to be produced again respectively at $t\sim100$~s and $t\sim$~hours by the $\beta$-decays of $^{56}$Ni and $^{56}$Co, with a half-life of ${\sim}6$ and ${\sim}77$~days, respectively. Correspondingly, $Y_{^{56}\textrm{Ni}}$ starts its exponential decay at $t\sim 6 \textrm{ days} \sim 5\times 10^5$~s. There are no significant qualitative differences between the two simulations.

\begin{figure}
  \centering
  \includegraphics[width=0.49\textwidth]{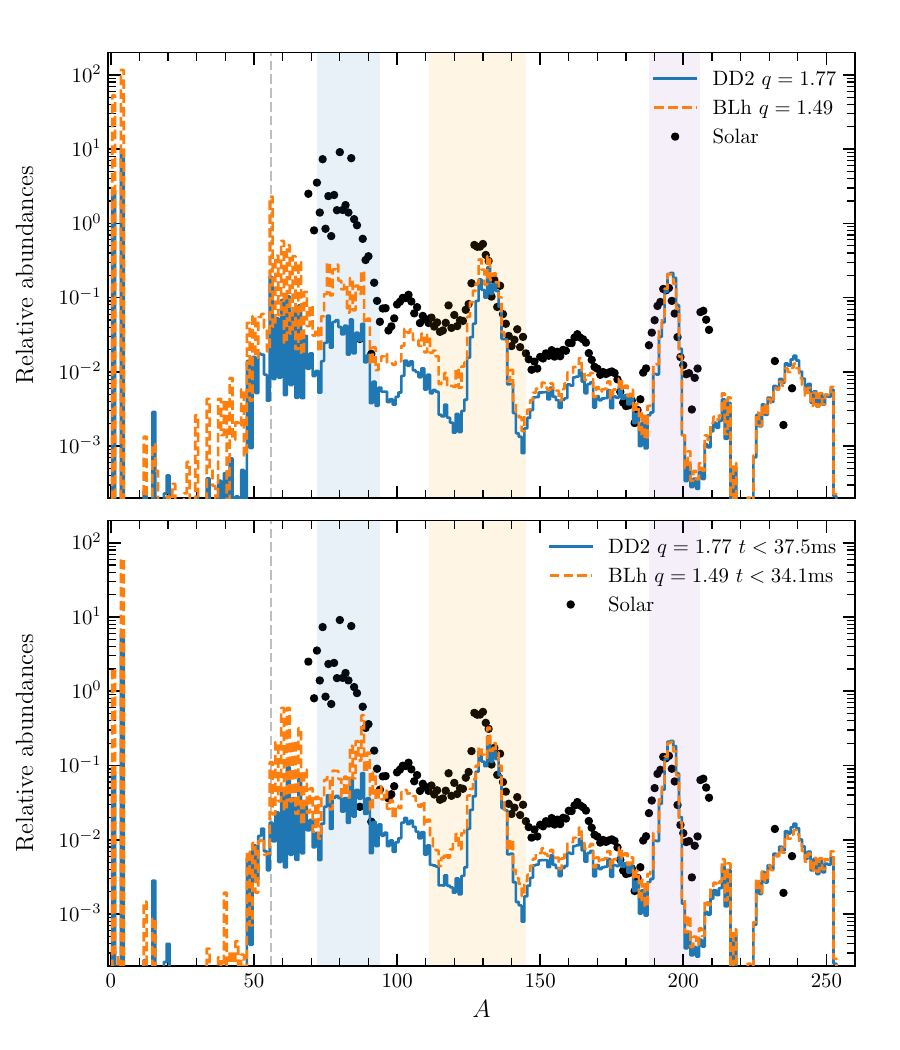}
  \caption{Nucleosynthesis yields for the DD2 $q=1.77$ (solid blue) and BLh $q=1.49$ (dashed orange lines) simulations at grid resolution SR as obtained from \knecnn. Top: results considering all times of ejection. Bottom: considering dynamical ejecta only. The vertical shaded regions highlight the position of the $r$-process peaks, while the vertical gray, dashed line shows the position of the $A = 56$ isotopes. Solar abundances are shown as reference.}
  \label{fig:nucleo}
\end{figure}

The final nucleosynthesis yields are shown in Fig.~\ref{fig:nucleo} for both binaries. Top and bottom panels show, respectively, the yields computed for the total ejecta and for the early ejecta. The abundances from the solar residual $r$-process from \citet{2020MNRAS.491.1832P} are reported for reference. Abundances are normalised by fixing the overall fraction of elements with $A \in [170,\,200]$ to be the same for all sets of abundances.

The nucleosynthesis yields comprise $r$-process and iron group elements.
Nuclei with $220 \lesssim A \lesssim 230$ are long-lived and are expected to $\alpha$-decay on timescales of $\gtrsim \mathcal{O}(10)$~years. The most abundant nuclei are free protons and $^4$He. Protons are already produced in high-entropy tails of the dynamical ejecta but the wind contributes to their abundance of about an order of magnitude. Similarly, $^4$He production is boosted of a factor three by the high-$Y_e$ winds, not associated to $r$-process nucleosynthesis. This is in agreement with the analysis of \cite{Perego:2020evn} for comparable-masses BNSM. The formation of light elements with $A<40$ in high-entropy regions of the ejecta is strongly suppressed, whereas in low entropy and low-$Y_e$ regions seed nuclei are formed already close to stability. 
$^{56}$Ni is mostly produced by the high $Y_e \gtrsim 0.45$ matter of the neutrino-driven winds originating at the edge of the remnant disk. Its subsequent decay originates $^{56}$Co and then $^{56}$Fe.
The second and third $r$-process peaks are instead fully produced by the early, low $Y_e \lesssim 0.22$ tidal ejecta.
The presence of a proton-rich neutrino-driven wind is consistent with what was found in the 2D (axisymmetric) Newtonian radiation-hydrodynamics simulations of \citet{Just:2023wtj}, which further evolved the dynamical ejecta from conformally flat Smooth Particle Hydrodynamics 3D simulations. Those simulations however do not find the same amount of $^{56}$Ni as found here, see also \citep{Jacobi:2025eak}. From their Tab.~1 and Fig.~3 we estimate ${\sim}3\times10^{-3}\Mo$ of proton-rich ejecta which is comparable to Tab.~I of \citet{Jacobi:2025eak} from our simulations.

\section{Multimessenger Observables}
\label{sec:obs}

\subsection{Kilonova light curves}
\label{sec:kn}

\begin{figure*}
  \centering
  \includegraphics[width=.99\textwidth]{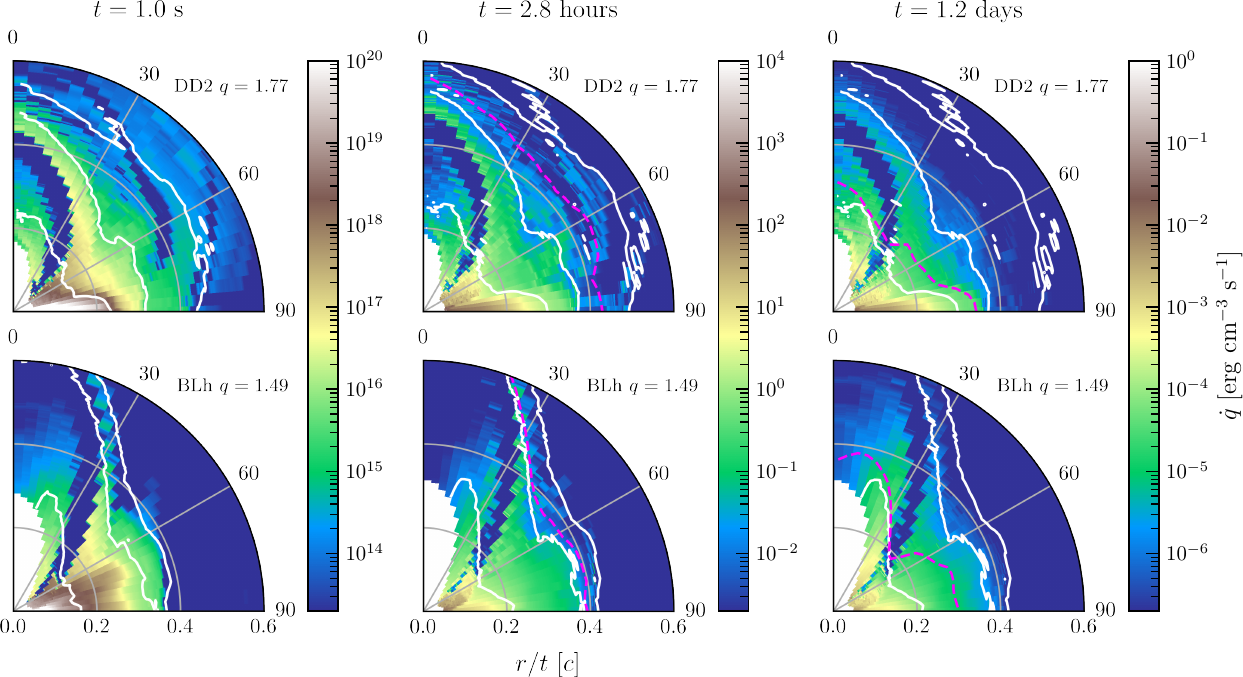}
  \caption{Local heating rate in the ejecta as function of the velocity coordinate $r/t$ and the polar angle at $t=1, 10^4$, and $10^5$\,s (\ie, ${\sim}1$~s, $2.8$~hours, and $1.2$~days.) White contours show contours at $\rho \left( \frac{t}{1\,\text{s}}\right)^{-3} = 0.01,\, 0.1,\, 5\,\gccm$. The dashed magenta lines show the position of the photosphere.}
  \label{fig:heating_rate}
\end{figure*}

\begin{figure*}
  \centering
  \includegraphics[width=.99\textwidth]{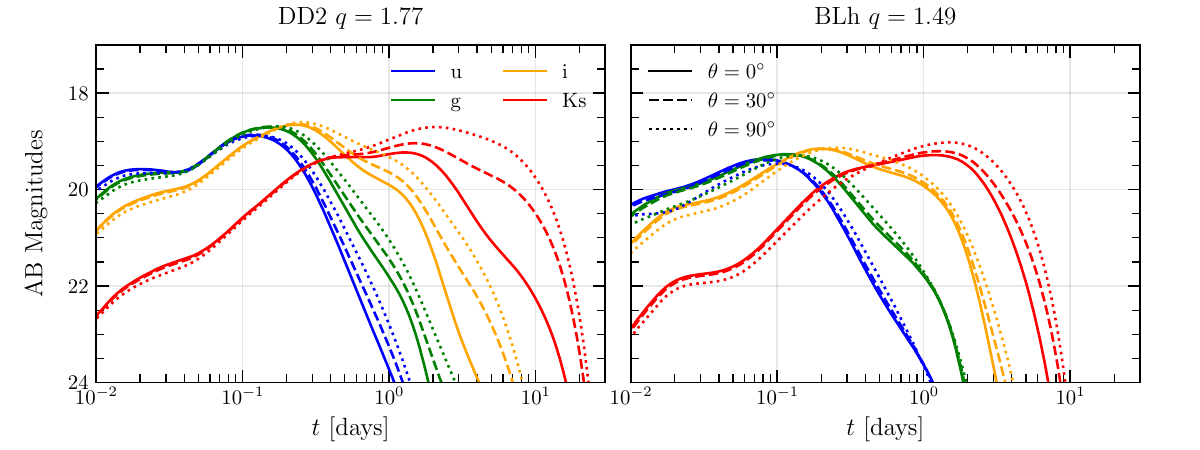}
  \caption{AB apparent magnitudes predicted by \knecnn for the Gemini bands $u$, $g$, $i$ and $K_s$ for an observer at a distance of $40$ Mpc and observation angles $\theta=0^{\circ}$ (face-on observer), $30^{\circ}$ and $90^{\circ}$ (edge-on observer).}
  \label{fig:kn}
\end{figure*}

Kilonovae shine with peak luminosities in the range ${\sim}10^{40-44}~\ergsec$ that are reached on day timescales after merger \citep{Metzger:2010sy}. Significant differences are expected in their properties (peak time, luminosities and colors) depending on the type of merging system. Roughly, a remnant NS can enhance the luminosity and the color variety due to the energy available from the central rotating object which is converted on longer timescales than those available in case of black hole formation by several mechanisms (\eg~spiral-waves). Such a variety in phenomenology makes detection strategies challenging but offers, at the same time, the opportunity to infer the source also in the absence of GWs or other counterparts.

Kilonova light curves are computed in \knecnn by assuming a multi-temperature black body emission model~\citep{Morozova:2015bla,Wu:2021ibi}. The model calculates the luminosity as $L_{\rm obs}=L_{\rm photo}+L_{\rm above}$, where the first term is the emission from the photosphere and the second is the contribution from the optically thin region outside the photosphere. The latter is given by the integrated heating rate $\dot{\epsilon}_{\rm nucl}$ from all the mass shells above the photosphere. The position of the photosphere is computed as the coordinate radius at which the optical depth reaches 2/3. The photosphere contribution dominates the light curves at early times and from most of the polar angles. The integral on heating rates is typically an order of magnitude smaller at early times and becomes the dominant contribution at $t\sim1-5$~days for mass shells at $\theta\sim0-60$~deg. In the equatorial plane, the photosphere emission dominates until $t\approx13$~days in BLh~$q=1.49$ and $t>30$~days in DD2~$q=1.77$.
Note that homologous expansion is reached on timescales of about $t\gtrsim0.5$~s postmerger for most of the ejecta, while the tidal ejecta typically accelerate by 0.01 - $0.05 c$ due to strong $r$-process heating up to $t \lesssim 20$~s.

Figure~\ref{fig:heating_rate} shows the energy deposition rate per unit volume $\dot{q}=\rho\,\dot{\epsilon}_{\rm nucl}$ in the ejecta as function of the velocity coordinate $r/t$ and the polar angle at three characteristic evolution times and for the two binaries, respectively. At early times $t\sim1$~s (left panels), energies of up to ${\sim}10^{20}\ergseccm3$ are deposited in the bulk of the ejecta ($\dot{\epsilon}_{\rm nucl}\sim10^{18}\ergsecg$, which has velocities ${\lesssim}0.2$~c (\cf~Fig.~\ref{fig:outflow_hist}). Consequently, the ejecta around the orbital plane $\theta\sim90^{\circ}$ are accelerated by nuclear heating on these timescales.
The outermost region above the remnant, $\theta\gtrsim45^{\circ}$, is populated by shocked heated ejecta.
In the DD2 $q=1.77$ binary, this matter is initially slightly neutron-rich and hosts weak $r$-process nucleosynthesis that generates a moderate heating rate at early times.
Further inside, the ejecta exhibit electron fractions around $Y_e\sim0.47$ and produce isotopes along the valey of stability so the heating rate is negligible in this region.
The innermost and more massive wind component is slightly proton-rich and the site for all the $^{56}$Ni production.
All the ejecta material is opaque to radiation and well inside the photosphere at this time.

At $t=2.8$~hours heating rates decreased by about eight orders of magnitudes. The bulk of ejecta is approximately at hundreds of million kilometers from the remnant. The photosphere has reached the outer region of the ejecta's bulk and it is shown in the figure as a dashed magenta line. Its shape is nonspherical and roughly follows the density contour of the ejecta at $\rho_{\rm photo}(t\sim1\,{\rm hr})\sim10^{-13}\,\gccm$. Note the significant differences in heating rates above the remnant bewteen the DD2 and BLh (central panels) and that the BLh photosphere is significantly less spherical than the DD2.

At $t\sim1$~day, heating rates in the bulk of the ejecta further decrease to $\dot{\epsilon}_{\rm nucl}\lesssim10^{10}\ergsecg$. The photosphere has penetrated further inside the expanding ejecta, $\rho_{\rm photo}(t\sim1{\rm day})\sim10^{-15}\gccm$, and reaches the innermost shell of the computational domain by ten days.

Kilonova light curves are shown in Fig.~\ref{fig:kn} and presented using the AB magnitude system, 
\begin{equation}
m_\mathrm{AB} = -2.5 \log_{10} \left(\frac{\int f_\nu(h\nu)^{-1} e(\nu) d\nu}
{\int{3631 \mathrm{Jy}(h \nu)^{-1} e(\nu) d\nu}}\right) \ , 
\end{equation}
where here $\nu$ is the light frequency, $f_\nu$ is the observed flux density at frequency $\nu$ from a distance of $40$ Mpc (see Eq.~(14) of \cite{Wu:2021ibi}) and $e(\nu)$ are filter functions for different Gemini bands. The light curves are calculated from \knecnn{} by removing a contribution from early-time (pre-merger) spurious ejecta, which is present in the $(3+1)$D simulations due to the atmosphere treatment and effects at the stellar surface. The main effect of such spurious ejecta on the light curves is to unphysically enhance the early blue peak (more below). 

The light curves peak at a few days in the (near-) infrared ($K_s$ in the Figure) band. The peak is mainly associated with the early-time, neutron-rich dynamical ejecta of tidal origin with late-time contributions from the winds. Consequently, these peaks are brighter for an observer placed edge-on ($\theta=90^{\circ}$) with respect to the initial orbital plane. Inspecting heating rates from \knecnn{} indicates that the slope of this late ``red'' kilonova is determined by the nuclear energy released by the combination of the decays of $r$-process elements and the Ni-Co-Fe chain~\citep{Jacobi:2025eak}.

Other luminosity peaks are found at ${\sim}10$~minutes and ${\sim}2-3$~hours, in the UV/optical ($u$ in the Figure) bands.
The early ``blue'' peak at ${\sim}2-3$~hours  is due to $r$-process decays at early times (${\sim}1$s) that heat up ejecta material in a wide angular region with density $\rho(t=1s)\lesssim0.1\gccm$, see left panels of Fig.~\ref{fig:heating_rate}, with $\dot{q}\sim10^{17}\ergsecg$. The fluid develops a large temperature gradient (around the blue-yellow region near the intermediate density isocontour in Fig.~\ref{fig:heating_rate}) which is initially inside the photosphere. At ${\sim}2.8$~days the photosphere crosses these layers and photospheric emission ($L_{\rm photo}$) produces the peak in the light curves.
The blue peak is the largest luminosity peak for the light curves of the DD2 binary (left panel). Compared to BLh, it originates from more ejecta mass and stronger $r$-process (see Sec.~\ref{sec:ejecta} and Fig.~\ref{fig:outflow_hist}.) The dependence on the viewing angle of the blue peak is weaker then that of the red peak. The blue peaks are only slightly enhanced for a edge-on ($\theta=90^{\circ}$) observer. This is due to the wide angular distribution of the low-opacity ejecta in Fig.~\ref{fig:outflow_hist} and the local heating rate plot in the central panels of Fig.~\ref{fig:heating_rate}.

The earlier blue peak at ${\sim}15$~minutes (in $u$ band) is due to $\beta$-decay of free neutrons and is the ``precursor'' already identified in previous works~\citep{Metzger:2014yda,Combi:2022nhg,Magistrelli:2024zmk}. Neutrons start to decay close to the photosphere; their mass fraction at the photosphere reaches up to $10\%$. Consequently, the precursor emission is due to both $L_{\rm photo}$ and, as the photosphere moves inwards, to the contribution from optically thin regions. We verified the origin of the precursor by performing further \knecnn simulations in which the heating rate associated to free neutrons is switched off; a summary of these experiments are presented in Appendix \ref{app:freen}. The precursor peaks at ${\sim}10$~minutes in the UV and is delayed and dimmer at longer optical wavelengths (\cf~Fig.~\ref{fig:lum_bol_free_n}). Free neutrons originates from the long-term evolution of the fast tails of the ejecta $v_\infty\sim0.4$~c, as already noted in the literature. They are distributed over the entire polar angle, \cf~Fig.~10 of \citep{Radice:2018pdn}.
In our asymmetric binaries, however, the matter producing decaying free neutrons is ejected before the moment of merger and constitute the dynamical tidal tails. A further difference from previous works is that these ejecta consist of mildly neutron-rich material $Y_e\approx0.4$. This is likely due to the use of M1 neutrino transport. An important caveat on these results is that the artificial atmosphere in the $(3+1)$D simulations and the mapping of the ejecta profile to \knecnn (symmetry reduction and Newtonian gravity) may impact the quantitative details of the computed precursor.

Finally, the fast tails of the dynamical ejecta are expected to generate synchrotron radiation as the ejecta remnant interacts with the surrounding interstellar medium (ISM), \eg~\citep{Nakar:2011cw,Hajela:2021faz}. As an illustration, we compute the radio light curves using the analytical model of \cite{Sadeh:2022enp}. The ejecta is modeled by a broken power law profile composed of a shallow bulk component and a steeper fast tail component,
\be
m_{\rm ej} = m_0\, %
\begin{cases}
  (x/x_0)^{-s_{\rm ft}} & x > x_0\\
  (x/x_0)^{-s_{\rm kN}} & 0.1 \leq x \leq x_0
\end{cases}\,,
\ee
where $x=\gamma(v_{\infty})v_{\infty}/c$ and $\gamma(v_{\infty})$ the Lorentz factor. Optimal fitting parameters for DD2 (BLh) binary are $x_0=0.201$, $m_0=7.69\times10^{-4}$, $s_{\rm ft}=4.24$ and $s_{\rm kN}=1$ ($x_0=0.444$, $m_0=1.15\times10^{-4}$, $s_{\rm ft}=6.97$ and $s_{\rm kN}=1$). The differences in these fits reflect the rather different velocity profiles of the binaries, see the bottom right panel of Fig.~\ref{fig:outflow_hist}. In particular, the DD2 profile has a less steep fast tail extending at larger velocities than the BLh.

\begin{figure}
  \centering
  \includegraphics[width=.49\textwidth]{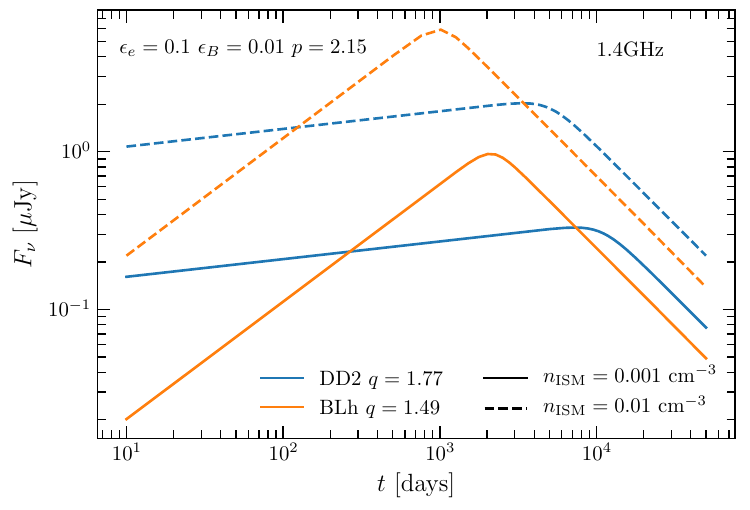}
  \caption{Radio light curve at $1.4$~GHz generated from the analytical model of \citet{Sadeh:2022enp}
    and the simulated profiles. The fiducial values of the model's parameter are $\epsilon_e=0.1$ and $\epsilon_B=0.1$ for the conversion efficiencies of the internal energy of the shock to the energy of the accelerated electrons and amplified magnetic field, $p=2.15$ the spectral index of the non-thermal electrons, and ISM density $n_{\rm ISM}=0.001, 0.01$~cm${}^{-3}$. The sources are placed at a distance of $40$~Mpc.}
  \label{fig:kna}
\end{figure}

Light curves at $1.4$~GHz are then computed from a 1D forward-reverse shock model of the ejecta, by using fiducial values of $\epsilon_e=0.1$ and $\epsilon_B=0.01$ for the conversion efficiencies of the internal energy of the shock to the energy of the accelerated electrons and amplified magnetic field, $p=2.15$ the spectral index of the non-thermal electrons, and ISM density $n_{\rm ISM}=0.001$~cm${}^{-3}$ (or the more optimistic $n_{\rm ISM}=0.01$~cm${}^{-2}$). The results are shown in Fig.~\ref{fig:kna}. The non-thermal flux increases in time until the reverse shock propagates to the bulk of the ejecta; after the peak the velocity is subrelativistic and the model approaches the Sedov-Taylor self-similar solution.

According to the employed model and using the conservative $n_{\rm ISM}=0.001$~cm${}^{-3}$ (solid lines), the BLh afterglow flux density peaks at ${\sim}6$ years postmerger ($F_{\nu\ \rm peak}\approx0.97\mu$ Jy). The DD2 afterglow is relatively dimmer ($F_{\nu\ \rm peak}\approx0.3\mu$Jy) and peaks at ${\sim}30$~years postmerger. A more optimistic value $n_{\rm ISM}=0.01$~cm${}^{-2}$ (dashed lines), increases the peak fluxes to a few $\mu$Jy and shifts them to earlier postmerger times of ${\sim}3$ and ${\sim}14$~years respectively for BLh and DD2. These signals might be detectable with radio facilities like VLA, Aperitif, ASKAP, MeerKAT, and SKA, in particular if the source is well localized by a coincident gravitational-wave detection~\eg~\citep{Hotokezaka:2016clu,Dobie:2021qya}.
Previous work on comparable masses BNSM identified the origin of these counterparts as due to the fast tails generated at the collisional shock between the stars~\citep{Metzger:2014yda,Hotokezaka:2018gmo,Radice:2018pdn}. Here, fast tails are instead mostly generated by tidal torque. Hence, mass asymmetry in combination with the EOS dependence further enriches the phenomenology of kN afterglows \citep{Nedora:2021eoj}.

\subsection{Gravitational waves}
\label{sec:gws}

Gravitational wave (GW) observations offer the unique opportunity to unambiguously identify the merger remnant and thus establish the connection between possible electromagnetic counterparts and their central engine. The detection of GWs from the remnant, in particular, appears possible with third-generation detectors~\citep{Breschi:2022ens}.

\begin{figure*}
  \centering
  \includegraphics[width=0.9\textwidth]{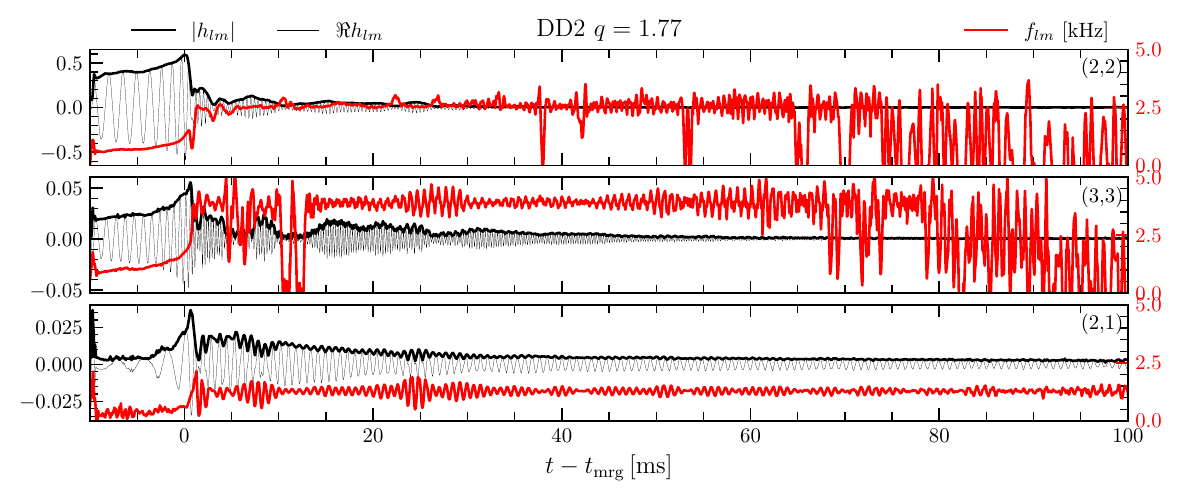}
  \caption{Multipolar gravitational waves strain for the simulation DD2 $q=1.77$ SR.
    The black lines show the amplitude and real part (in geometric units $G=c=\Mo=1$) of the dominant GW modes with multipolar indexes $\lm=22,33,21$. The red line shows the modes' frequency. At times $t-t_{\rm mrg}\gtrsim40$~ms the 21 mode has an amplitude larger than the 33 and comparable to the 22.}
  \label{fig:gws:DD2}
\end{figure*}

\begin{figure*}
  \centering
  \includegraphics[width=0.9\textwidth]{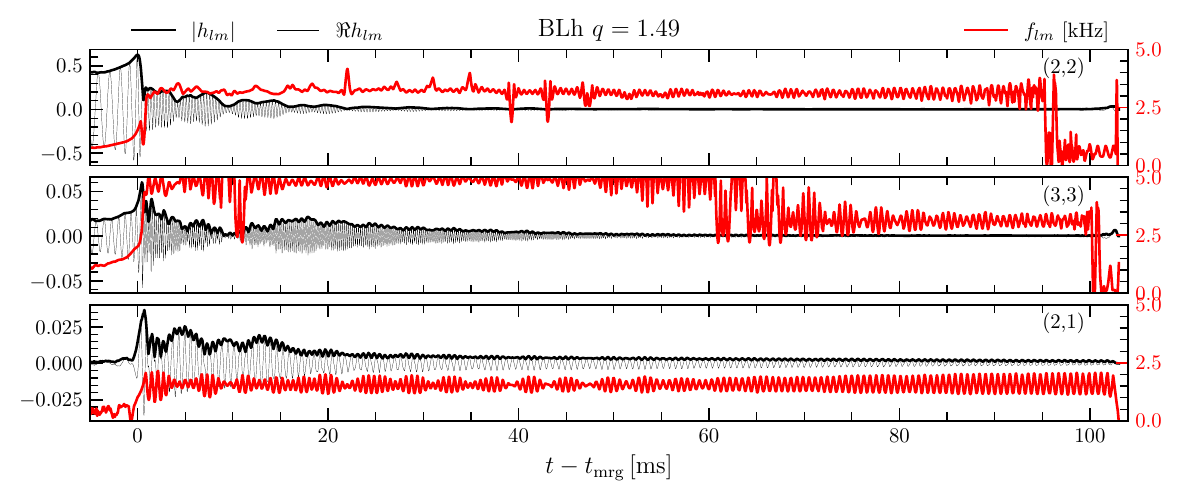}
  \caption{Same as Fig.~\ref{fig:gws:DD2} but for the  BLh $q=1.49$ SR K2 simulation.}
  \label{fig:gws:BLh}
\end{figure*}

Figure~\ref{fig:gws:DD2} and~\ref{fig:gws:BLh} show the three dominant multipoles of the GWs radiated by the two binaries. In each figure, from top to bottom, the black lines show the amplitude and real part~\footnote{We use geometric units $G=c=\Mo=1$ for the GW strain and recall that
$h:=h_+-ih_{\times}=D_{\rm L}^{-1}\sum_\lm h_{\lm}(t){}^{-2}Y_{\lm}(\iota,\psi)$, where the extrinsic properties of the source are incorporated in the luminosity distance $D_{\rm L}$ and via the spin-weighted spherical harmonics ${-2}^{}Y_{\lm}$ (sky position).}
of the $\lm=22,33,21$ multipoles. The red line shows the modes' frequency. The postmerger signal is characterized at early times by the 22 emission from the NS remnant at a peak frequency of $f_{22}\simeq2.5$~kHz for the DD2 binary and $f_{22}\simeq3.5$~kHz for the BLh binary. The latter frequency is higher because the remnant is more compact. Other significant emission channels are the 33 mode with an amplitude of about an order of magnitude smaller than the 22 and a frequency $f_{33}\sim 3/2 f_{22}$, and the 21 mode with an even smaller amplitude but a lower frequency $f_{21}\sim 1/2 f_{22}$.

Most of the GW energy is radiated up to times $t-t_{\rm mrg}\lesssim20$~ms \citep{Bernuzzi:2015opx}. After these times the merger dynamics are driven by the viscous processes discussed above rather than gravitational-wave backreaction. During the viscous phase, the one-armed ($m=1$) spiral motion of the remnant generates a weak but persistent in time signal, \eg~\citep{Paschalidis:2015mla,East:2015vix,Radice:2016gym}. Indeed, a close inspection of the GW amplitude at $t-t_{\rm mrg}\gtrsim40$~ms reveals that the 21 mode has an amplitude larger than the 33 and comparable to the 22. Note also that the 22 and 33 GW frequencies evolve to lower values at later times and acquire a progressively richer frequency content. By contrast, the 21 frequency remains steady over the entire simulated time.

The GW $m=1$ mode is a strong signature for the production of a remnant NS and a potential smoking gun for the identification of tidal disruption mergers like those simulated here. The detection of such mode could also convey information on the remnant's extreme matter as, for example, QCD phase transitions \citep{Espino:2023llj}.
The detectability of the $m=1$ mode by third-generation GW experiments is favoured by the frequency $f_{21}\sim1$~kHz, relatively lower with respect to other postmerger frequencies, but heavily suppressed by the small amplitude of the signal and its monochromatic nature. In the context of equal-mass quasi-circular mergers, \citet{Radice:2016gym} found that detection would be possible only with an optimally oriented source at $10$~Mpc. For the binary considered here similar results apply, thus making unlikely a postmerger detection of these remnants signals. However, a third-generation detection of the lower frequency signal from the inspiral-merger is expected to deliver accurately the mass and mass ratio. This implies immediately a high probability for the merger to produce an electromagnetic counterpart if a large mass \textit{and} a large mass ratio ($q\gtrsim1.5$) are detected because those binary parameters produce bright electromagnetic signals also in case of a rapid collapse to black hole \citep{Bernuzzi:2020txg}.

\section{Conclusion}
\label{sec:conclusion}

We discussed 3D general-relativistic radiation-hydrodynamics simulations of asymmetric BNSMs in which the secondary NS undergoes a (partial) tidal disruption. For both a stiff and a soft microphysical EOS, the remnant NS is stable on a timescale comparable to its cooling time and it is surrounded by massive, neutron-rich discs. The tidal disruption generates neutron-rich, $Y_e\sim0.1$, ejecta at merger. During the subsequent evolution, the remnant’s spiral waves and neutrino heating generate winds of mass ${\gtrsim}10^{-2}\Mo$ with a broad composition, $Y_e\sim0.1-0.55$, that varies with the line of sight to the binary orbital plane, see Fig.~\ref{fig:ye2d:DD2}-\ref{fig:ye2d:BLh}. The wind components are persistent over the cooling timescale of the remnant; they decelerate as the spiral-arms dynamics weakens and are driven by neutrino irradiation by the end of the simulations.
The simulations highlight several new features that further enrich BNSM observational prospects.

The nucleosynthesis yields reveal, for the first time in these ab-initio simulations, the production of ${}^{56}$Ni and the subsequent decay chain to $^{56}$Co and $^{56}$Fe. These elements are produced by the $Y_e\sim0.5$ material of the neutrino-driven winds \citep{Perego:2014fma,Martin:2015hxa,Perego:2017fho}. A follow-up paper has reported an in-depth analysis of possible observational signatures~\citep{Jacobi:2025eak}. These can includes signatures in the light curves, spectra as well as $\gamma$-rays associated with the Ni-Co-Fe radioactive decays~\citep{2020ApJ...889..168K} (compare to \eg~\citep{Churazov:2014bga,Jerkstrand:2020hlf} for similar results in type Ia and core-collapse supernovae.)

We stress that the large range of $Y_e$ found in the winds is a common prediction of M1 neutrino transport schemes in the BNSM problem \citep{Foucart:2016rxm,Vincent:2019kor,Radice:2021jtw,Kiuchi:2022nin,Radice:2023zlw}. Systematic studies of the impact of neutrino transport with more sophisticated schemes will further quantify the robustness of these results \citep{Foucart:2020qjb,Radice:2021jtw,Zappa:2022rpd,Foucart:2024npn,Cheong:2024buu}. Another example of the impact of neutrino transport schemes on the ejecta modeling is presented in Appendix~\ref{app:M01}, where ejecta properties discussed in this paper are compared with those obtained with a M0 transport scheme.

Our synthetic kilonova light curves are characterized by a main (near-) infrared peak at a few days postmerger and a UV/optical peak at a few hours postmerger, see Fig.~\ref{fig:kn}. The temporally extended ``red'' component is qualitatively similar to what found for asymmetric BNSM that undergo accretion-induced collapse \citep{Bernuzzi:2020txg}.
The early ``blue'' peak is instead a generic feature related to presence of a fast and high-$Y_e$ component from both the dynamical ejecta and the wind. Such ejecta component is driven by neutrino irradiation.
In-line with previous comparable-masses simulations, we identify a UV/optical precursor on a timescale of tens of minutes due to the $\beta$-decay of free neutrons. For the considered binaries, free neutrons are produced in fast ejecta tails with mildly neutron-rich material ($Y_e \sim 0.4$) and are distributed at all latitudes. This is different from previous results, which report the precusor originates from neutron rich ejecta component $Y_e\sim0.22$, but used a more approximate neutrino transport scheme. More details and caveats on the precuros origin and interpretation are discussed in Appendix~\ref{app:freen}.


Our 2D ray-by-ray simulations show that the photosphere has a rather asymmetric shape, Fig.~\ref{fig:heating_rate}. This suggests that (full) 2D and 3D radiation-hydrodynamics simulations might be very important to obtain precise light curves. Work in this direction has started by various groups \citep{Tanaka:2013ana,Nativi:2020moj,Collins:2022ocl,Shingles:2023kua}; although the inclusion of hydrodynamics and complete ejecta data remains a challenge for future studies \citep{Magistrelli:2024zmk}. Work is also ongoing to explore the impact of frequency-dependent (vs. gray) and composition-informed opacities~\eg~\citep{Tanaka:2013ana,Wollaeger:2017ahm}, which may impact light curves on the days timescales (Magistrelli {\it et al.} 2025, In prep.).

The presented simulations indicate that dynamical ejecta of tidal origin can have fast tails reaching $v_\infty\sim0.6-0.8$~c. Such tails can generate a non-thermal kN afterglow that can be detectable years after merger in radio and X-band~\citep{Nakar:2011cw,Hajela:2021faz}. While previous work identified these counterparts as the result of shocked-heated dynamical ejecta, \eg~\citep{Radice:2018pdn}, here we showed that they can be produced also by tidal disruption. Further, the quantitative details of the outflows between the two considered models are actually rather different. On the one hand, this highlights that emissions are highly degenerate with binary parameters. On the other hand, this enriches the kN landscape and motivates further studies.

Gravitational waves are characterized by a signature in the $m=1$ channels due to the one-armed non-axisymmetric dynamics of the remnant \citep{Paschalidis:2015mla,East:2015vix,Radice:2016gym}. The dominant frequency, $f_{21}\sim1$~kHz, is lower than the merger frequency and persists over time as a quasi-monochromatic postmerger signal. While this is a potential smoking gun for the identification of these tidal disruption mergers, the detectability by third-generation detectors is hindered by the small amplitude and might be possible only for optimal source orientations and nearby events~\citep{Radice:2016gym}.

\section*{Acknowledgments}

The authors thank Almudena Arcones, Federico Guercilena, Oliver Just and Giacomo Ricigliano for interactions and comments. The authors thank Alejandra Gonzalez for preparing the waveform release on the CoRe database.
SB thanks Prof.~David Hilditch for important suggestions. 

SB and MJ knowledge support by the EU Horizon under ERC Consolidator Grant, no. InspiReM-101043372. FM acknowledges support from the Deutsche Forschungsgemeinschaft (DFG) under Grant No.406116891 within the Research Training Group RTG 2522/1.
DR acknowledges support from the Sloan Foundation, from the U.S.~Department of Energy, Office of Science, Division of Nuclear Physics under Award Number(s) DE-SC0021177 and DE-SC0024388, and from the National Science Foundation under Grants No.~PHY-2011725, PHY-2020275, AST-2108467, PHY-2116686, and PHY-2407681.
The work of AP is partially funded by the European Union - Next Generation EU, Mission 4 Component 2 - CUP E53D23002090006 (PRIN 2022 Prot. No. 2022KX2Z3B).

Simulations were performed on SuperMUC-NG at the Leibniz-Rechenzentrum (LRZ) Munich and and on the national HPE Apollo Hawk at the High Performance Computing Center Stuttgart (HLRS).
The authors acknowledge the Gauss Centre for Supercomputing e.V. (\url{www.gauss-centre.eu}) for funding this project by providing computing time on the GCS Supercomputer SuperMUC-NG at LRZ (allocations {\tt pn36ge}, {\tt pn36jo} and {\tt pn68wi}). The authors acknowledge HLRS for funding this project by providing access to the supercomputer HPE Apollo Hawk under the grant number INTRHYGUE/44215 and MAGNETIST/44288.
Postprocessing and development runs were performed on the ARA cluster at Friedrich Schiller University Jena. The ARA cluster is funded in part by DFG grants INST 275/334-1 FUGG and INST 275/363-1 FUGG, and ERC Starting Grant, grant agreement no. BinGraSp-714626.

\section*{Data Availability}

Data generated for this study will be made available upon reasonable request to the corresponding authors.
Gravitational waveforms are available at the \href{http://www.computational-relativity.org/}{CoRe} database~\citep{Gonzalez:2022prs} as records \coredbentry{THC}{0108} (BLh K2), \coredbentry{THC}{0109} (BLh K1) and \coredbentry{THC}{0110} (DD2).


\bibliographystyle{mnras}


\appendix

\section{Free-neutrons decay and blue kilonova precursor}
\label{app:freen}

\begin{figure}
  \centering
    \includegraphics[width=.49\textwidth]{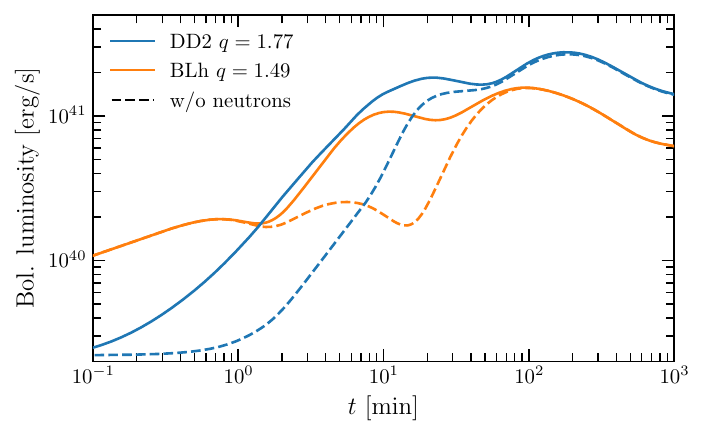}
  \caption{Bolometric lightcurves for both binaries on minutes
    timscales. Dashed lines show the lightcurve from a simulation in
    which the heating from free neutrons is explicitly removed.} 
  \label{fig:lum_bol_free_n}
\end{figure}

\begin{figure}
  \centering
    \includegraphics[width=.49\textwidth]{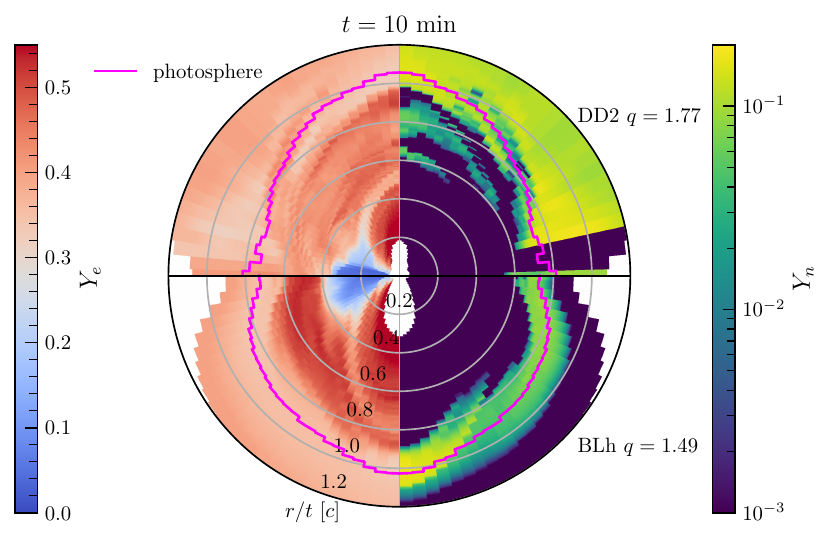}
  \caption{Distribution of the initial electron fraction (left) and neutron abundance at 10~minutes (right) in the ejecta for both binaries (top and bottom respectively). The magenta line marks the position of the photosphere.}
  \label{fig:ejecta_no_n}
\end{figure}

This appendix summarizes some investigations on the origin of the blue
precursor in the light curves of Fig.~\ref{fig:kn}.
The $\beta$-decay of free neutrons in the outermost shells of the
dynamical ejecta have been proposed as the origin of a blue precursor
peaking on timescales of tens of minutes postmerger~\citep{Metzger:2014yda}. 
Such a precursor has been identified by \citet{Combi:2022nhg} in
neutron-rich $Y_e\sim0.22$ dynamical ejecta fast tails from 
equal-mass BNSM simulations employing GRMHD, the M0 neutrino scheme
of~\citet{Radice:2018pdn} and a similar setup to ours for the
synthetic light curves. 
\citet{Magistrelli:2024zmk} confirmed this result and clearly
assessed, with similar experiments as those reported in this appendix,
that free-neutron decay powers almost entirely the blue peak in their
model. The numerical relativity data used by
\citet{Magistrelli:2024zmk} refer to an unequal-mass BNSM simulation
with the M0 neutrino scheme; the remnant is short-lived and
gravitational collapse happens on dynamical timescales. 

Bolometric light curves on minutes timescales are shown in
Figure~\ref{fig:lum_bol_free_n} for both binaries.
Dashed lines indicate the lightcurve from simulations where the
heating from free-neutron decay is explicitly removed. In both cases the
blue precursor is significantly reduced, thus supporting the
interpretation that the main origin is free-neutron decay.

We further analyze free-neutrons abundances in the \knecnn profiles. 
Figure~\ref{fig:ejecta_no_n} shows the electron (left) and neutron (right)
fraction in the ejecta at $t\sim10$~minutes for both binaries (top and
bottom respectively). 
The figure demonstrates that free neutrons fractions $Y_n\sim0.1$ are
present at all latitudes in the mildly neutron-rich ejecta, $Y_e\sim0.4$. 
This is different from previous simulations. A possible reason for the
discrepancy is the use of M1 (vs. M0) neutrino transport that enhances the electron
fraction of the ejecta (see Appendix~\ref{app:M01}). 
Further, the mass asymmetry of the binary and the tidal disruption
dynamics may lead to faster neutron-rich dynamical ejecta tails (than
in equal-mass binaries) and thus inhibit neutron capture. 

Free neutrons originates from the fast tail $v_\infty \sim0.4$, see
Fig.~\ref{fig:outflow_hist}.
Figure~\ref{fig:ejecta_no_n} shows that during the \knecnn evolution
the ejecta velocity is significantly boosted and, in the 
Newtonian \knecnn setup, can reach velocities $v\approx c$. 
The symmetry reduction employed to map the 3D
ejecta profile to 2D and the subsequent \knecnn evolution ray-by-ray
are likely responsible for such high ejecta velocity because the (averaged)
fluid elements are  forced to move radially. Moreover, the artificial
atmosphere employed in the 3D simulations may quantitative impact
these results because free neutrons originates from low density tidal
tails. 
More work is needed to clarify these aspects, although the presence
of a UV precursor powered by free-neutron decay appears a qualitative
robust feature in different binaries and simulations with different physics
(eventually with quantitative differences).

\section{M1 vs. M0 neutrino transport}
\label{app:M01}

\begin{figure}
  \centering
  \includegraphics[width=.5\textwidth]{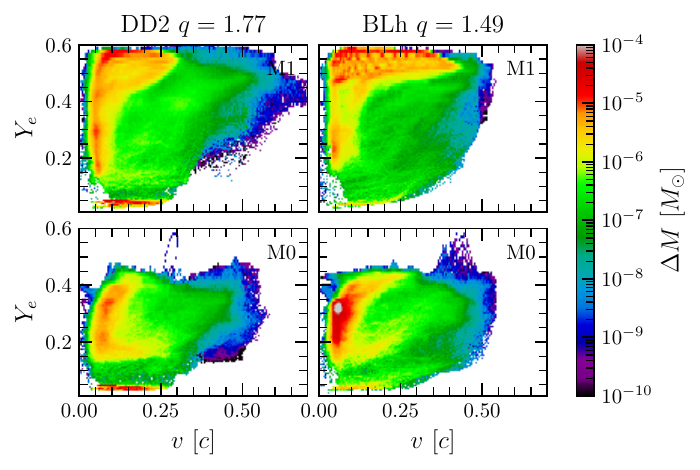}
  \caption{Comparison of the ejecta properties for the BLh $q=1.49$ and DD2 $q=1.77$ runs (upper panels) and comparable runs with the M0 neutrino transport scheme (lower panels). The panels show 2D mass-weighted histograms of ejecta's electron fraction and velocity. M1 transport produces more proton-rich ejecta; notably the electron fraction is pushed to larger values than those obtained with M0 neutrino transport. Ejecta velocities with the M1 transport are higher (lower) for proton (neutron) rich matter than those with the M0 transport.}
  \label{fig:ye_vel_hist}
\end{figure}

This appendix discusses the impact of the neutrino transport scheme on the ejecta properties. To this aim, we compare our new M1 simulations with very similar simulations performed with the M0 neutron transport scheme of~\citet{Radice:2018pdn}; both schemes are implemented in the same \thc~code. The M0 simulation settings are practically identical to those presented here; the K1 turbulent viscosity scheme is used for both binaries.

Figure~\ref{fig:ye_vel_hist} shows 2D mass ejecta histograms of the ejecta's $Y_e$ and velocity. The figure demonstrates significant differences in the ejecta velocity and composition resulting from the two different transport schemes. The ejecta mass spans a broader range in $Y_e$ in the M1 simulations. Ejecta velocities in the M1 runs are comparable to or higher than those in the M0 runs in the proton-rich material. However, significantly less mass with high velocity and low-$Y_e$ is found: the M1 transport scheme suppresses these ejecta components with respect to the M0 (lower-right areas in the panels).

These results agree with the detailed study presented by \citet{Zappa:2022rpd}. The latter reference concluded that nucleosynthesis yields are robust provided that both neutrino emission and absorption are simulated (either with M0 or M1), but it considered relatively shorter simulations than those presented here. The longer simulations presented here allow us to better asses the impact of the neutrino-driven wind on the nucleosynthesis. For the current study, the M1 scheme is key to obtain ejecta with $Y_e\gtrsim0.45$ and thus for the nucleosynthesis of light elements with $A\sim56$.
In relation to the study in Appendix~\ref{app:freen}, we note that the neutrino transport scheme can also significantly impact on the free-neutron decay contribution to the kN because the free-neutron contribution is associated with high-velocity and high $Y_e\sim0.4$ ejecta.

\bsp	
\label{lastpage}
\end{document}